\documentclass[journal,twoside]{IEEEtran}
\IEEEoverridecommandlockouts
\usepackage{amsmath,amssymb,amsfonts}
\usepackage{algorithmic}
\usepackage{graphicx}
\usepackage{textcomp}
\usepackage{xcolor}
\usepackage{gensymb}
\usepackage{multirow, multicol}

\usepackage[font=footnotesize]{caption}
\usepackage[font=footnotesize]{subcaption}
\usepackage{graphicx} 
\usepackage{makecell}
\usepackage[detect-all]{siunitx}
\usepackage[acronym]{glossaries-prefix}
\usepackage{cite}

\usepackage{array}
\usepackage{booktabs}
\usepackage{flushend}
\newcommand{\mycomment}[1]{}
\usepackage{bm}
\usepackage{etoolbox}
\makeatletter

\newacronym[plural={LWAs},prefixfirst={a\ }, prefix={an\ }]{lwa}{LWA}{leaky wave antenna}
\newacronym{osb}{OSB}{open stopband}
\newacronym{lh}{LH}{left-handed}
\newacronym{crlh}{CRLH}{composite right left-handed}
\newacronym{siw}{SIW}{substrate integrated waveguide}
\newacronym{tem}{TEM}{transverse electromagnetic}
\newacronym{vna}{VNA}{Vector Network Analyzer}

\date{ }

\def\BibTeX{{\rm B\kern-.05em{\sc i\kern-.025em b}\kern-.08em
    T\kern-.1667em\lower.7ex\hbox{E}\kern-.125emX}}
\begin{document}

\title{Meandering Microstrip Leaky-Wave Antenna with~Dual-band Linear–Circular Polarization and~Suppressed Open Stopband}

\author{Pratik~Vadher,~\IEEEmembership{Student Member,~IEEE,}
Giulia~Sacco,~\IEEEmembership{Member,~IEEE,}
and~Denys~Nikolayev,~\IEEEmembership{Senior Member,~IEEE}%

\thanks{Manuscript received April 25, 2023, revised October 3, 2023.}%
\thanks{This project has received funding from the French Agence Nationale de la Recherche (ANR) through the Project MedWave under Grant ANR-21-CE19-0045, the European Union's Horizon 2020 research and innovation programme under grant agreement N$^\circ$899546 (Marie Sk\l odowska-Curie REACH-IT project), and the European Union's Horizon Europe research and innovation programme under grant agreement N$^\circ$101063966 (Marie Sk\l odowska-Curie IN-SIGHT project). \textit{(Corresponding author: Denys Nikolayev, denys.nikolayev@deniq.com)}}%
\thanks{P. Vadher, G. Sacco, and D. Nikolayev are with the Univ Rennes, CNRS, IETR -- UMR 6164, FR-35000 Rennes, France.}%
}%


\maketitle

\begin{abstract}
    This paper proposes a dual-band frequency scanning meandering microstrip leaky-wave antenna with linear polarization in the Ku-band and circular polarization in the K-band. This is achieved by making use of two spatial harmonics for radiation. The unit cell of the periodic microstrip antenna contains three meanders with mitred corners. To ensure circular polarization, a theoretical formulation is developed taking into account the delay caused by the microstrip length intervals. It defines the unit cell geometry by determining the length of the meanders to ensure that axial ratio remains below \SI{3}{\dB} throughout the operational band. Moreover, the meanders are used to provide better control over scanning rate (the ratio of change of angle of maximum radiation with frequency) and reduce spurious radiation of harmonics by ensuring single harmonic operation within the operational band. To guarantee continuous scanning through broadside direction, open stopband is suppressed using mitered angles. The antenna is designed on a 0.254-mm substrate making it suitable for conformal applications. The fabricated antenna shows a backward to forward beam steering range of \SI{72}{\degree} (\SI{-42}{\degree} to \SI{30}{\degree}) in the K-band (\SIrange[range-units=single, range-phrase=--]{19.4}{27.5}{\GHz}) with circular polarization and of 75$\degree$ (--15$\degree$ to  60$\degree$) in the Ku-band (11–15.5~GHz) with linear polarization.
\end{abstract}

\glsresetall

\section{Introduction}
Periodic \glspl{lwa} are a class of travelling-wave antennas that radiate energy at the discontinuities of the guiding medium \cite{walter_traveling_1965}. Changes in frequency causes dispersion within the guiding medium, resulting in varying excitation phases at the discontinuities. This, in turn, alters the main beam pointing direction in the radiation pattern of the antenna with respect to frequency. One-dimensional periodic \glspl{lwa} \mbox{(1D \glspl{lwa})} typically radiate a fan-beam in the E-plane. The direction of maximum radiation changes in the H-plane with the change in frequency \cite{balanis_antenna_2015}.

Many 1D periodic \glspl{lwa} have been proposed using different guiding media. \Glspl{lwa} proposed in \cite{agarwal_multilayered_2021, sarkar_60_2020, sabahi_compact_2018, caloz_crlh_2008, liu_substrate_2012} use \gls{siw} or half-mode \gls{siw} to support the travelling wave and usually employ a TE$_\mathrm{n0}$ or in certain cases TM$_{11}$ mode \cite{carrara_tm11_2023} for radiation. However, to reduce fabrication complexity, lower manufacturing costs, and to enable the creation of compact antennas, microstrip-based guiding media \glspl{lwa} are an appealing solution \cite{sacco_analysis_2021, zhao_circularly_2022, yang_full-space_2010, duan_transversal_2021,wang_periodic_2022}. Meandering microstrip \glspl{lwa} radiate due to the discontinuities created at the edges which results in a net magnetic current responsible for radiation \cite{james_microstrip_1986, wang_periodic_2022}.

Since \glspl{lwa} are periodic structures, infinite spatial harmonics exist in the guiding media \cite{volakis_antenna_2007, balanis_antenna_2015}. Many \gls{siw} based \glspl{lwa} that use higher order Floquet modes for radiation have been proposed recently \cite{rahimi_higher-order_2021, guan_scanning_2018}. Although the previous works report scanning due to higher spatial harmonics, the scanning is due to two or three spatial harmonics at the same time, which leads to more than one beam radiated by the antenna. However, it is desirable to have a single beam scanning operation over a high scanning range. This is possible when only a single spatial harmonic is responsible for radiation. Hence, better separation of different spatial harmonics is necessary.

To be suitable for on-body conformal applications \cite{soh_-body_2012, soh_wearable_2015, herssensSurveyOnBodyAntenna2023, soares_analysis_2023}, the antenna must be flexible and bendable~\cite{tianConformalPropagationNearomnidirectional2020}. For such a purpose, antennas based on microstrip technology are an excellent candidate. Additionally, the mechanical properties such as tensile modulus of the dielectric substrate also play an important factor. Hence, the Rogers 3003 substrate ($\varepsilon_r~=~$3.0) is preferred in this work due to its low tensile modulus (\SI{823}{\mega\pascal}) making it flexible.

Moreover, it is desirable to have circular polarization for many different applications where the alignment of the receiving and transmitting antenna may impact the overall performance of the system such as radars \cite{hasch_millimeter-wave_2012}, satellite communications and on-body antenna system \cite{soares_analysis_2023,matthews_development_2009, vadher_-body_2022}. 

Presence of \glspl{osb} also impacts the scanning of the antenna through the broadside direction \cite{henry_broadside-scanning_2016, jackson_leaky-wave_2008}. Several techniques have been proposed to suppress or completely mitigate \glspl{osb} \cite{otto_circular_2014, rahmani_backward_2017, paulotto_novel_2009}. At the \gls{osb} frequency, the input impedance matching is poor and the Bloch impedance ($Z_\mathrm{s}$) has a high imaginary value. Hence, the aim is to minimize the high imaginary value of impedance to reduce the effects of \gls{osb}. The idea of beam scanning due to higher spatial harmonic, limited to a single-band and linear polarization, has been presented in \cite{vadher_higher_2023} with no suppression of \gls{osb}.

\begin{figure}[!t]\centerline{\includegraphics[width=0.5\textwidth]{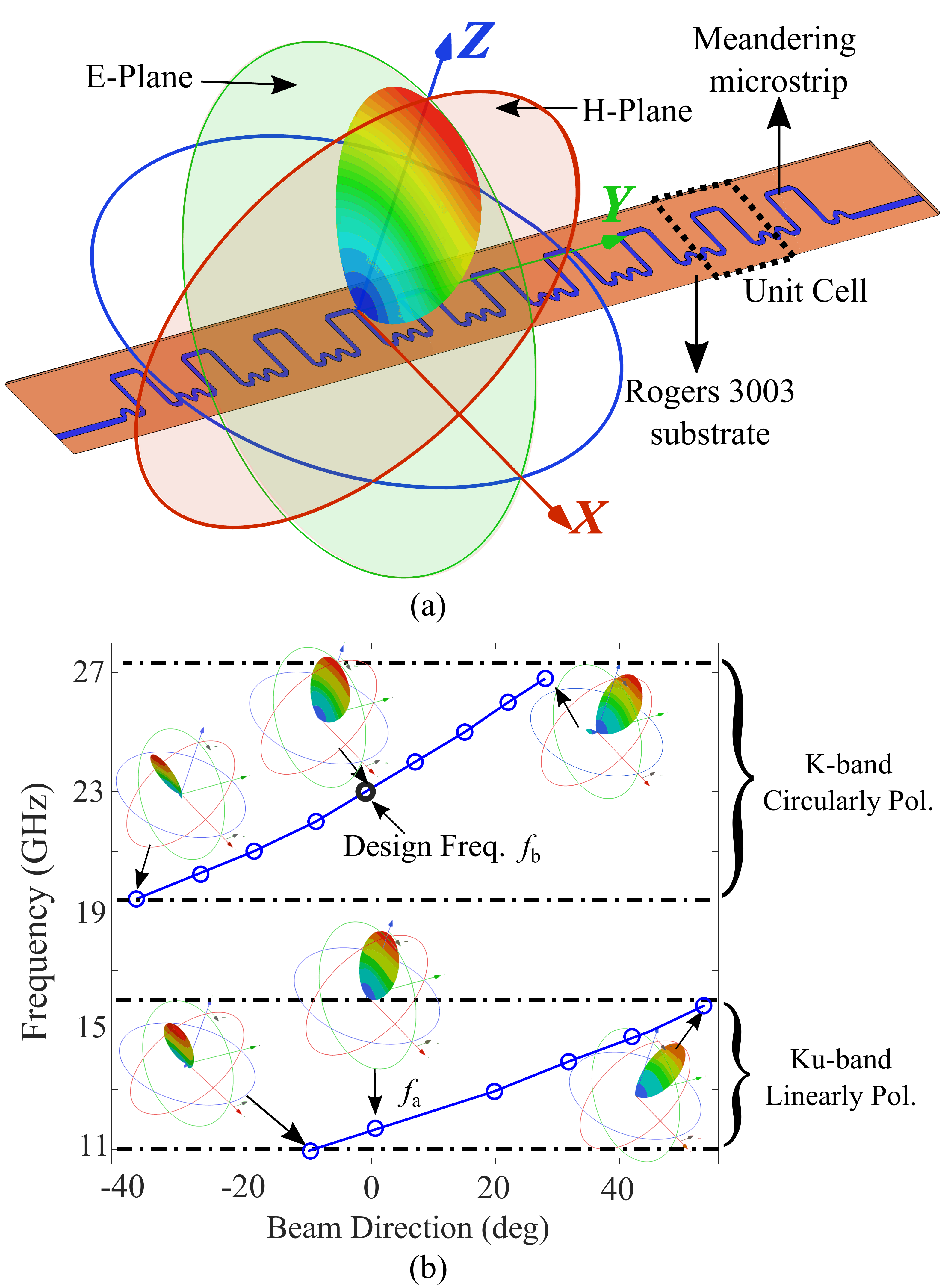}}
\caption{(a) Configuration of the proposed 10 unit cells periodic frequency scanning antenna with radiation pattern shown in the broadside direction. The meandering microstrip is etched on top layer while bottom layer is copper. (b)~Beam scanning operation of the proposed LWA as a function of frequency in the Ku-band {and K-band}. The antenna radiates a fan-beam in the E-plane (X--Z plane)\textcolor{blue}{,} while in the H-plane (Y--Z plane) the antenna changes the direction of maximum radiation with change in frequency.}
\label{thetavsfreq}
\end{figure}

In this paper, single-layer PCB meandering microstrip based \gls{lwa} is proposed with dual-band operation without the use of vias shown in Fig~\ref{thetavsfreq}(a). The Ku-band operation of the antenna is due to the $n = -1$ spatial harmonic and it exhibits linear polarization, while the second band of operation at K-band is due to $n = -2$ and it depicts circular polarization as described in Fig.~\ref{thetavsfreq}(b). The unit cell of the periodic antenna consists of three meanders to ensure single-beam operation by providing better separation between the different spatial harmonics.

The structure of the paper is as follows. Section \ref{sec:radiationmechanism} details the concept of higher spatial order in a microstrip-based unit cell followed by theoretical formulation for the dimensions required to have circular polarization for a unit cell with single meander. In Section \ref{sec:3meanders}, the unit cell is modified by adding two smaller meanders to increase the separation of spatial harmonics and ameliorate performance of circular polarization over the larger frequency operation range. The smaller meanders also result in better control over scanning rate (i.e. the ratio between the change in the angle of maximum radiation and the change in frequency ${\Delta\theta}/{\Delta f}$). A technique to remove \gls{osb} is discussed using Bloch impedance subsequently. Section \ref{sec:measurement} contains the comparison of simulations and measurements for the final design. In Section \ref{sec:conclusion}, the conclusions of the study are presented.

\begin{figure}[!t]\centerline{\includegraphics[width=0.45\textwidth]{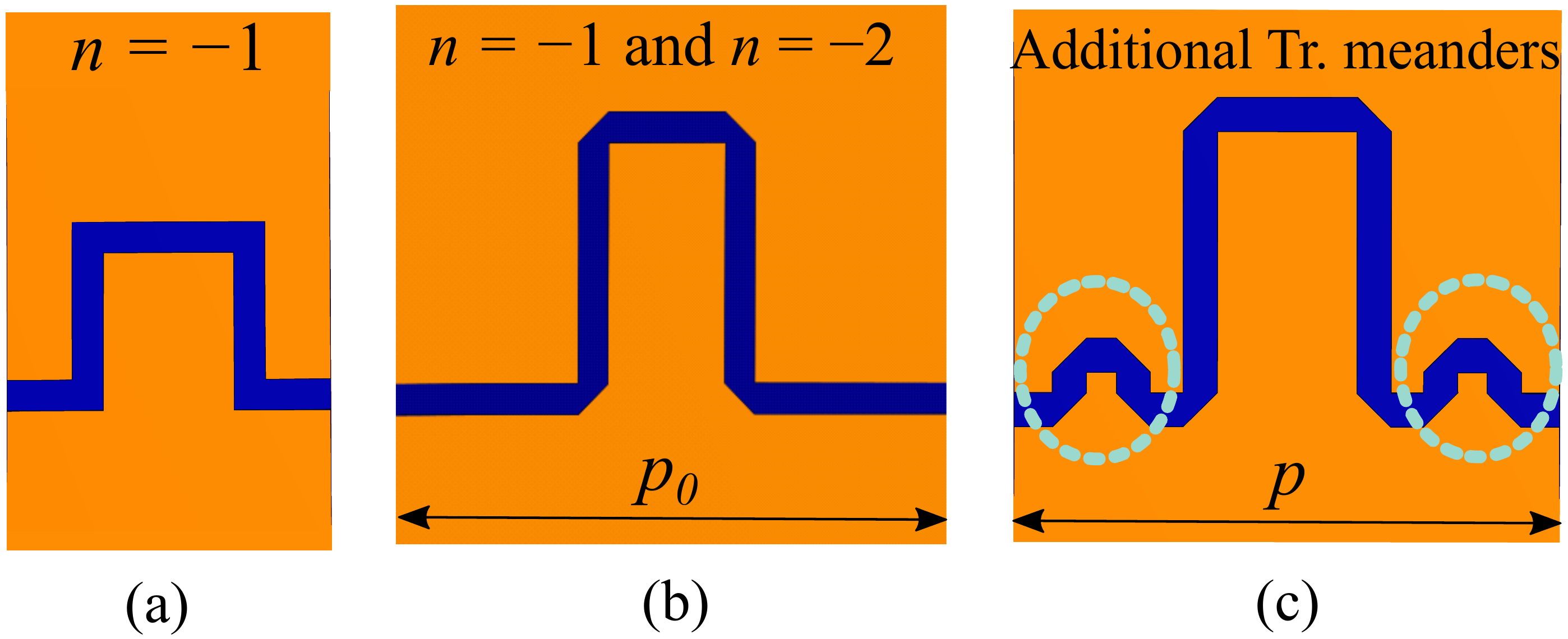}}
\caption{Evolution towards final design of proposed unit cell: (a)~conventional unit cell operating in single band and radiating due to the $n=-1$ spatial harmonic, (b)~dual band operating unit cell radiating due to $n=-1$ and $n=-2$ spatial harmonics, and (c)~modified unit cell with the introduction of two meanders ($p<p_0$) to improve the scanning range and the circular polarization performance.}
\label{cellevol}
\end{figure}

\section{Radiation Mechanism and Principle for~Antenna Design}
\label{sec:radiationmechanism}
Fig.~\ref{thetavsfreq}(a) shows the design of the proposed microstrip-based \gls{lwa} with mitred corners. As detailed in previous works on microstrip based \glspl{lwa} \cite{wang_periodic_2022}, the radiation occurs due to the magnetic current at the corners of the meandering microstrip. 

The evolution towards the final design of the unit cell is shown in Fig.~\ref{cellevol}. Conventional microstrip-based \gls{lwa} \cite{ cheng_approximate_2019, wang_periodic_2022, chen_circular-polarized_2019, wood_curved_1979} operate in the radiation zone due to spatial harmonic of $n=-1$ like the unit cell shown in Fig.~\ref{cellevol}(a). This unit cell can be modified to operate in the $n=-2$ spatial harmonic, by increasing the pathlength at the desired operational frequency. By introducing mitred corners and properly choosing the length of the interconnecting microstrip lines between the corners the antennas radiates in circular polarization (see Fig.~\ref{cellevol}(b)). The unit cell operates in the radiation zone associated with $n=-1$ and $n=-2$ spatial harmonics.

To improve the separation of radiation zone due to two harmonics ($n=-1,-2$) we propose the geometry in Fig.~\ref{cellevol}(c). The unit cell is modified with two additional meanders resulting in improvement in scanning range and circular polarization. The antenna has been optimized for operation in the circularly polarized K-band using the spatial harmonic of $n=-2$, and all theoretical formulations for the unit cell dimensions have been performed to ensure optimal performance in this frequency range.

\subsection{Higher order spatial harmonics in the unit cell based on~microstrip design}
\begin{figure*}[!t]
\centerline{\includegraphics[width=0.95\textwidth]{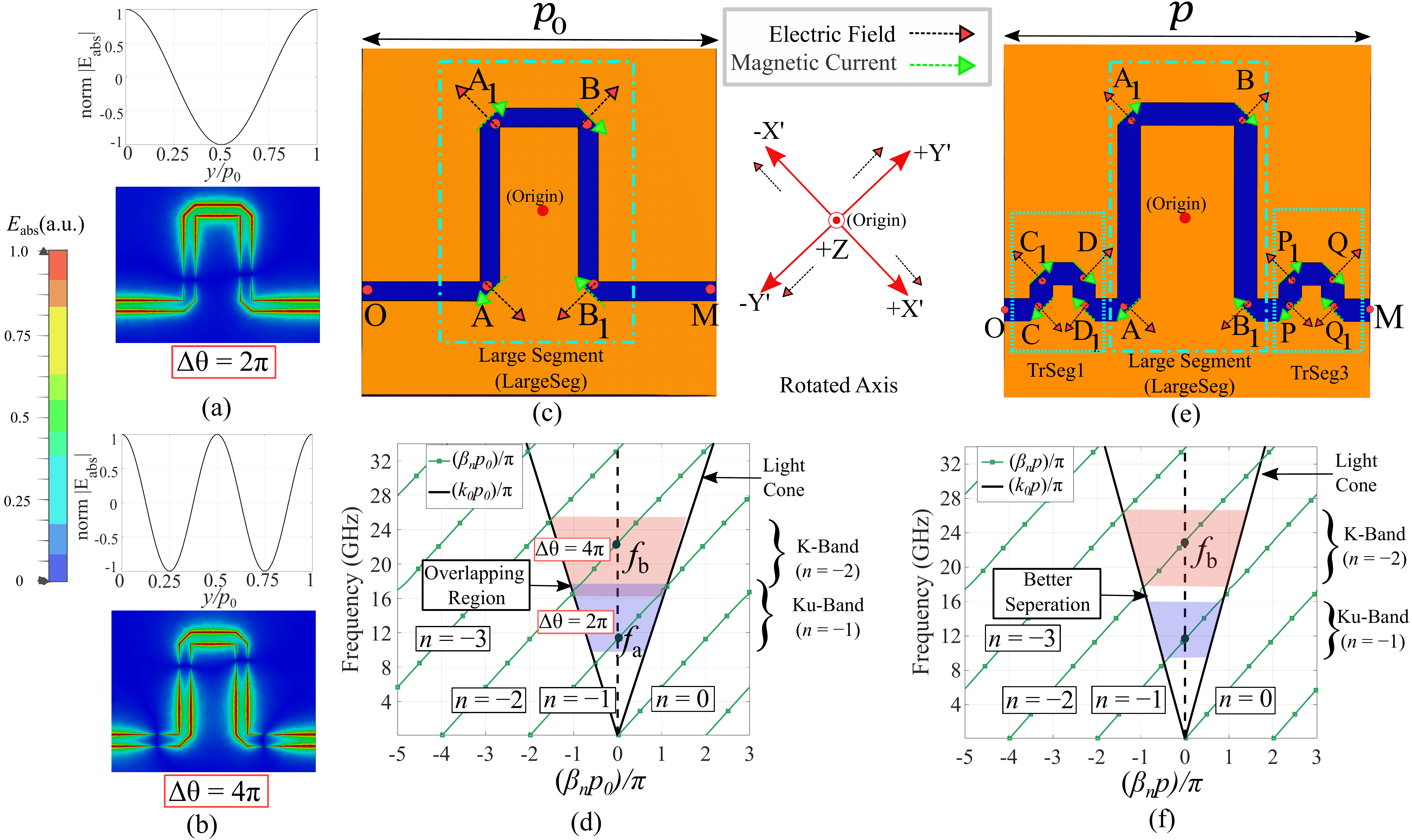}}
\caption{$E_{\mathrm{abs}}$ plots at phase \SI{0}{\degree} for the frequency at which broadside radiation occurs for (a)~$n=-1$ and (b)~$n=-2$. The sinusoidal graphs above $E_\mathrm{abs}$ plots are representative of the electric field variation across microstrip line throughout the unit cell. (c)~Unit cell with single large meander. Radiation sources for a single large meander are depicted in the figure. The four radiating magnetic current sources are shown at points $\mathrm{A}$, $\mathrm{A_1}$, $\mathrm{B}$ and $\mathrm{B_1}$. (d)~Brillouin diagram for the unit cell with the single meander. (e)~Unit cell design with two additional meanders that results in 12 radiation sources. (f)~Brillouin diagram for the improved unit cell with two additional meanders.}
\label{E_abs_ver2}
\end{figure*}

Fig.~\ref{E_abs_ver2}(c) shows the microstrip based unit cell with single meander with four radiating elements due to four mitred corners. Due to the periodicity ($p_0$) of the structure, infinite number of space harmonics exist due to Bloch-Floquet theorem \cite{jackson_leaky-wave_2011, collin_antenna_1969}. The phase constant of the $n^{\mathrm{th}}$ space harmonic $\beta_n$ satisfies
\begin{equation} 
\beta_{n}p_0 = \beta_{0}p_0+2 n\pi \label{eq1}
\end{equation}
where $n$ ranges from $-\infty$ to $+\infty$. Here $\beta_0$ is the zeroth order spatial harmonic of periodic \gls{lwa}

Fig.~\ref{E_abs_ver2}(d) shows the corresponding Brillouin diagram for the unit cell. The $n^{\mathrm{th}}$ spatial harmonic inside the light cone ($|\beta_n|<|k_0|$), results in radiation \cite{volakis_antenna_2007}. It is to be noted that multiple spatial harmonics can exist within the light cone resulting in multiple beam radiation \cite{rahimi_higher-order_2021}. 

The phase difference across the unit cell for the first two higher harmonics $n=-1$ and $n=~-2$, can be described from equation \eqref{eq1} with respect to fundamental harmonic.

Consequently, the direction of maximum radiation $\theta_{n}$ corresponding to $n^{\mathrm{th}}$ spatial harmonic is given by \cite{collin_antenna_1969, ishimaru_electromagnetic_2017, rahimi_higher-order_2021}
\begin{equation} 
\theta_{n}=\sin^{-1}(\beta_{-n}p_0/k_0p_0) \label{eq4}
\end{equation}
where $k_0$ is the free space wave number.

According to equation \eqref{eq4}, when the phase difference across the unit cell ($\beta_{-n}p_0$) is zero, the direction of main beam of the $n^{\mathrm{th}}$ spatial harmonic is in the broadside direction. Hence, at the frequencies of broadside radiation for spatial harmonics $n=-1$ and $n=-2$, the phase difference across the unit cell with periodicity $p_0$ from equation \eqref{eq1} is: 
\begin{subequations} 
\begin{align}
\beta_{-1}p_{0|f=f_\mathrm{a}} &= 0 \rightarrow \beta_{0}p_{0|f=f_\mathrm{a}} = 2 \pi \label{eq5} \\
\beta_{-2}p_{0|f=f_\mathrm{b}} &= 0 \rightarrow \beta_{0}p_{0|f=f_\mathrm{b}} = 4 \pi\,, \label{eq6}
\end{align}
\end{subequations}
where $f_{\mathrm{a}}$ and $f_{\mathrm{b}}$ are the frequencies corresponding to the broadside radiation for $n=-1$ and $n=-2$, respectively.
Fig.~\ref{E_abs_ver2}(a--b) depicts the variation of electric field magnitude at the two frequencies confirming the above two equations. To summarize, the first Ku-band radiation is due to spatial harmonic $n = -1$, while the second K-band radiation is due to spatial harmonic $n=-2$. 

Eq.~\eqref{eq6} implies that at $f = f_\mathrm{b}$, the total phase difference ($\phi_{p\mathrm{_0}}$) across the unit cell is equal to $4\pi$. Therefore, the equation to have broadside radiation for second harmonic at the design frequency $f=f_\mathrm{b}$ is, 
\begin{equation}
\begin{split}
    4\pi = \phi_{\mathrm{OA}_1|f=f_\mathrm{b}} +&\phi_{\mathrm{AA}_1|f=f_\mathrm{b}}+\phi_{\mathrm{A_1B}|f=f_\mathrm{b}}+ \\ &\phi_{\mathrm{BB_1}|f=f_\mathrm{b}}+\phi_{\mathrm{B_1{M}}|f=f_\mathrm{b}}  \label{eq7}
\end{split}
\end{equation}

Here, $\phi_{ij}$ is the phase difference across the line segment between the points $i$ and $j$, where $i, j \in \{\mathrm{A, A_1, B, B_1, O, M}\}$.
As indicated earlier, the reason for designing the antenna for the second harmonic is to obtain dual band operation due to spatial harmonics $n=-1$ and $n=-2$.

\subsection{Analysis for circular polarization in the radiation zone for spatial harmonic $n=-2$}

\begin{figure}[!t]
\centerline{\includegraphics[width=0.45\textwidth]{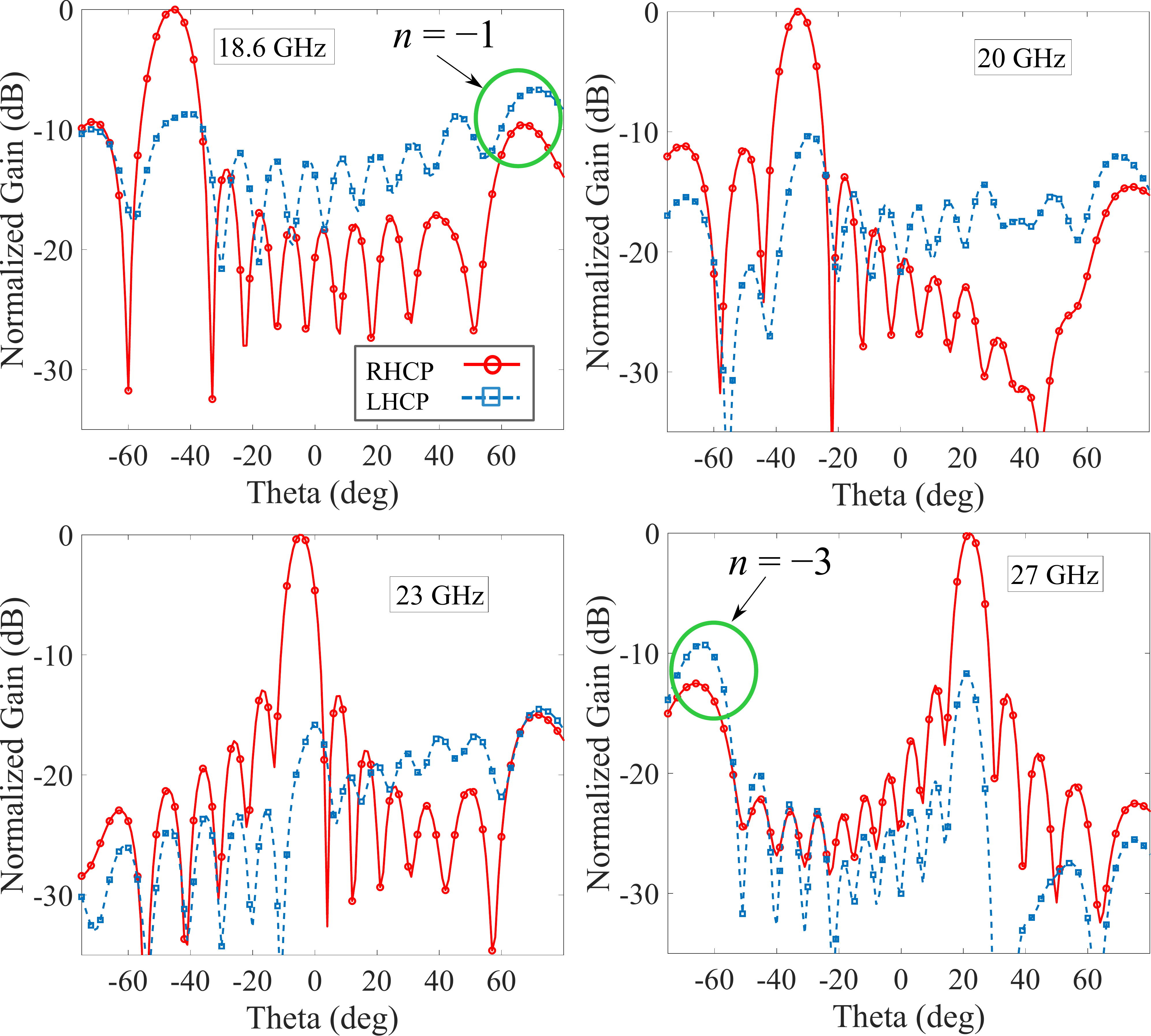}}
\caption{Normalized RHCP and LHCP gain in the H-plane obtained from simulation for the \gls{lwa} formed by connecting 10 unit cells with singular meander. The main lobe is due to spatial harmonic of $n=-2$. The realized gain of the main lobe varies from \SI{8}{\dB} to \SI{14}{\dB} across the frequency range.}
\label{RHCPLHCPGain_SingleFing}
\end{figure}

To analyse the circular polarization easily, the axis is rotated equal to the angle of mitred corners as shown in Fig.\ref{E_abs_ver2}(c). Hence there are two transverse components ($E_{\mathrm{X'}}$ and $E_{\mathrm{Y'}}$) radiating from the mitred corners. 

The ratio of the two transverse fields $E_{\mathrm{X'}}$ and $E_{\mathrm{Y'}}$, in the $\mathrm{X'}$ and $\mathrm{Y'}$ direction, respectively\cite{james_microstrip_1986}:

\begin{equation}
    \frac{E_{\mathrm{Y}'}}{E_{\mathrm{X}'}} = \mathrm{e}^{j(\phi_{\mathrm{AA_1}}+\phi_{\mathrm{A_1B}})}
    \label{EyEx}
\end{equation}

To have the circular polarization, the two transverse components of electric field ($E_\mathrm{X'}$ and $E_\mathrm{Y'}$) need to have a phase difference of odd multiple of $\pi/2$ \cite{balanis_antenna_2015}. Consequently, the condition to obtain circular polarization for a travelling-wave meandering microstrip antenna is stated by \cite{james_microstrip_1986}. \mycomment{Wood \textit{et al.}}

\begin{equation}
    \phi_{\mathrm{AA_1}}+\phi_{\mathrm{A_1B}} = (2k+1)\pi/2
    \label{EyEx_condition}
\end{equation}
where $k = 1, 2,..$

Since the desire is to operate in spatial harmonic $n=-2$, from equation \eqref{eq6}, the conclusion can be made to select $k=1$, hence
\begin{equation}
    \phi_{\mathrm{AA_1}}+\phi_{\mathrm{A_1B}} = 3\pi/2 
    \label{phidiffAB}
\end{equation}
\mycomment{Taking the ratio of equations \eqref{eq_Ex} and \eqref{eq_Ey}:}
\mycomment{The magnitude difference between the electric field ($\mathrm{E_{abs}}$) caused by the four radiating edges of a unit cell is negligible and therefore not taken into account in the analysis. Hence, 
\begin{equation}
    \begin{split}
    E_{X} &= \mathrm{E_{abs}}\times(\mathrm{e}^{-j\times\phi_\mathrm{A}}-\mathrm{e}^{-j\times\phi_{\mathrm{A_1}}})\\
    &=\mathrm{E_{abs}}\times(\mathrm{e}^{-j\times\beta_{0}l_{\mathrm{OA}}}-\mathrm{e}^{-j\times\beta_{0}(l_{\mathrm{OA}}+l_{\mathrm{AA_1}})})  \end{split}
    \label{eq_Ex}
\end{equation}
\mycomment{
\begin{equation}
\begin{split}
    E_{Y} &= |E_{abs}|\times(\mathrm{e}^{-j\times\phi_B}-\mathrm{e}^{-j\times\phi_{B1}})\\
    &= |E_{abs}|\times & (\mathrm{e}^{-j\times\beta_{0}(l_{OA}+l_{AA_1}+l_{A_1B})}-\\
    &\mathrm{e}^{-j\times\beta_{0}(l_{OA}+l_{AA_1}+l_{A_1B}+l_{BB_1})})
\end{split}
\end{equation}}
\begin{equation}
\begin{split}
    E_{Y} = \mathrm{E_{abs}}\times & (\mathrm{e}^{-j\times\beta_{0}(l_{OA}+l_{AA_1}+l_{A_1B})}-\\
    &\mathrm{e}^{-j\times\beta_{0}(l_{OA}+l_{AA_1}+l_{A_1B}+l_{BB_1})})
\end{split}
\label{eq_Ey}
\end{equation}
Here, $\phi_i$ represents the phase at point $i (= \mathrm{A, A_1, B, B_1})$, $\phi_{ij}$ represents the phase difference between the points $i$ and $j$ where $i, j \in \{\mathrm{A, A_1, B, B_1, O, M}\}$. }

Selecting $k\ge2$ would lead to similar result with equation~\eqref{EyEx} still being satisfied, however this would increase the overall period of the unit cell resulting in non desirable harmonics in the radiation zone.

Equation \eqref{phidiffAB} imposes a criterion on the the phase difference and hence the length of microstrip interval between two corners $\mathrm{A}$ and $\mathrm{B}$ for achieving circular polarization. Additionally, the length of the line segment between the points $i$ and $j$, indicated earlier, is represented by  $l_{ij}$. To have accurate length of line intervals, the model from \cite{sacco_analysis_2021} is chosen for analysis. For initial dimensions, at the design frequency $f=f_\mathrm{b}$, $l_{\mathrm{AA_1}}$ is taken such that $\phi_{\mathrm{AA_1}}=\pi$. This implies that $\phi_{\mathrm{A_1B}} = \pi/2$. 

From Fig.~\ref{E_abs_ver2}(d), $l_{\mathrm{AA_1}} = l_{\mathrm{BB_1}}$ which implies $\phi_{\mathrm{AA_1}}~=\phi_{\mathrm{BB_1}}$. Also, the unit cell is considered symmetric, hence $l_{\mathrm{OA}} = l_{\mathrm{B_1M}}$.

The phase difference across the large meander (consisting of $\mathrm{AA_1}$, $\mathrm{A_1B}$ and $\mathrm{BB_1}$) is defined as $\phi_{\mathrm{LargeSeg}}$. Therefore, from equation \eqref{phidiffAB}, in order to have the best circular polarization performance at the desired design frequency of $f = f_{\mathrm{b}}$, 
\begin{equation}
\begin{split}
    \phi_{\mathrm{LargeSeg}|f = f_{\mathrm{b}}} &= \phi_{\mathrm{AA_1}|f = f_{\mathrm{b}}}+\phi_{\mathrm{A_1B}|f = f_{\mathrm{b}}}+\phi_{\mathrm{BB_1}|f = f_{\mathrm{b}}} \\ &  =5\pi/2
    \label{importantconstraint}
\end{split}
\end{equation}
Consequently, there are two design constraints on the length of microstrip line intervals: 1) Equation~\eqref{eq7}, to construct a unit cell that has broadside radiation at $f=f_\mathrm{b}$. and 2) Equation~\eqref{importantconstraint}, to construct a unit cell with best circular polarization performance at $f=f_{\mathrm{b}}$. 

Table~\ref{SingleUnitCellDimensions_Table} depicts the dimensions of the unit cell based on the theory discussed at $f_\mathrm{b}=$ \SI{23}{\GHz}. The dimensions are mentioned in the form of $\lambda_\mathrm{b}$, corresponding wavelength to frequency $f_\mathrm{b}$ in the microstrip medium.

The dielectric is chosen as Rogers 3003 ($\varepsilon_\mathrm{r} = 3.0$) while the height of the substrate is $\mathrm{h_{sub}}=$ \SI{0.254}{\milli\meter}. The width of the microstrip $\mathrm{t_{50}}$ is equal to \SI{0.5}{\milli\meter}. It is important to note that in the microstrip medium the effective $\varepsilon_{\mathrm{r,eff}}(f=0) $ is $2.375$ calculated from \cite{pozar_microwave_2011}. To account for the dispersive nature of microstrip media, the model proposed by Pramanick and Bhartia from \cite{sadiku_comparison_2004} is considered to calculate $\varepsilon_{\mathrm{r,eff}}(f=f_\mathrm{b})$. This parameters are taken into consideration when calculating $\lambda_\mathrm{b}$ in the microstrip medium.

\captionsetup[table]{name=TABLE,labelsep=newline,textfont=sc, justification=centering}
\renewcommand{\arraystretch}{1.3}
\begin{table}
\begin{center}
\caption{Dimensions of the unit cell designed to operate in the $n=-2$ spatial harmonic and have circularly polarized fields. Here note that $\lambda_\mathrm{b}$ is the corresponding wavelength in the microstrip medium at $f=f_\mathrm{b}$} \label{SingleUnitCellDimensions_Table}
\begin{tabular}[t]{p{0.17\linewidth}>{\centering}p{0.17\linewidth}>{\centering\arraybackslash}p{0.27\linewidth}>{\centering\arraybackslash}p{0.18\linewidth}}
\toprule
    \multirow{2}*{\thead*{Line\\interval}}
    & \multirow{2}*{\thead*{Expressed \\ in $\lambda_\mathrm{b}$}}
    & \multirow{2}*{\thead*{Phase diff. across \\ line interval}} 
    & \multirow{2}*{\thead*{Physical \\ dim. (mm)}}\\ \\
\midrule
$\mathrm{OA}$& $3\lambda_\mathrm{b}/8$&$3\pi/4$ &3.21\\
$\mathrm{AA_1}$& $\lambda_\mathrm{b}/2$&$\pi$ & 4.44\\
$\mathrm{A_1B}$ & $\lambda_\mathrm{b}/4$&$\pi/2$ & 2.47\\
$\mathrm{BB_1}$ & $\lambda_\mathrm{b}/2$&$\pi$ & 4.44\\
$\mathrm{B_1M}$ & $3\lambda_\mathrm{b}/8$&$3\pi/4$ &3.21\\
$p_0$ &$2\lambda_\mathrm{b}$& $4\pi$ &5.8\\
$\mathrm{t}_{50}$ & - & - &0.5\\
$\mathrm{h}_{\mathrm{sub}}$ & - & - &0.254\\
\bottomrule
\end{tabular}
\end{center}
\end{table}

Ten such unit cells with single meander are cascaded next to each other to form a \gls{lwa} and full-wave simulations are performed in CST and ANSYS HFSS. Fig.~\ref{RHCPLHCPGain_SingleFing} shows the radiation pattern indicating the beam scanning with frequency and circular polarization nature of the antenna. As can be seen, at lower end of frequency range $f=$ \SI{18.6}{\GHz}, there is another beam of spatial harmonic of $n=-1$ limiting the scanning.

\begin{figure}[!t]
\centerline{\includegraphics[width=0.37\textwidth]{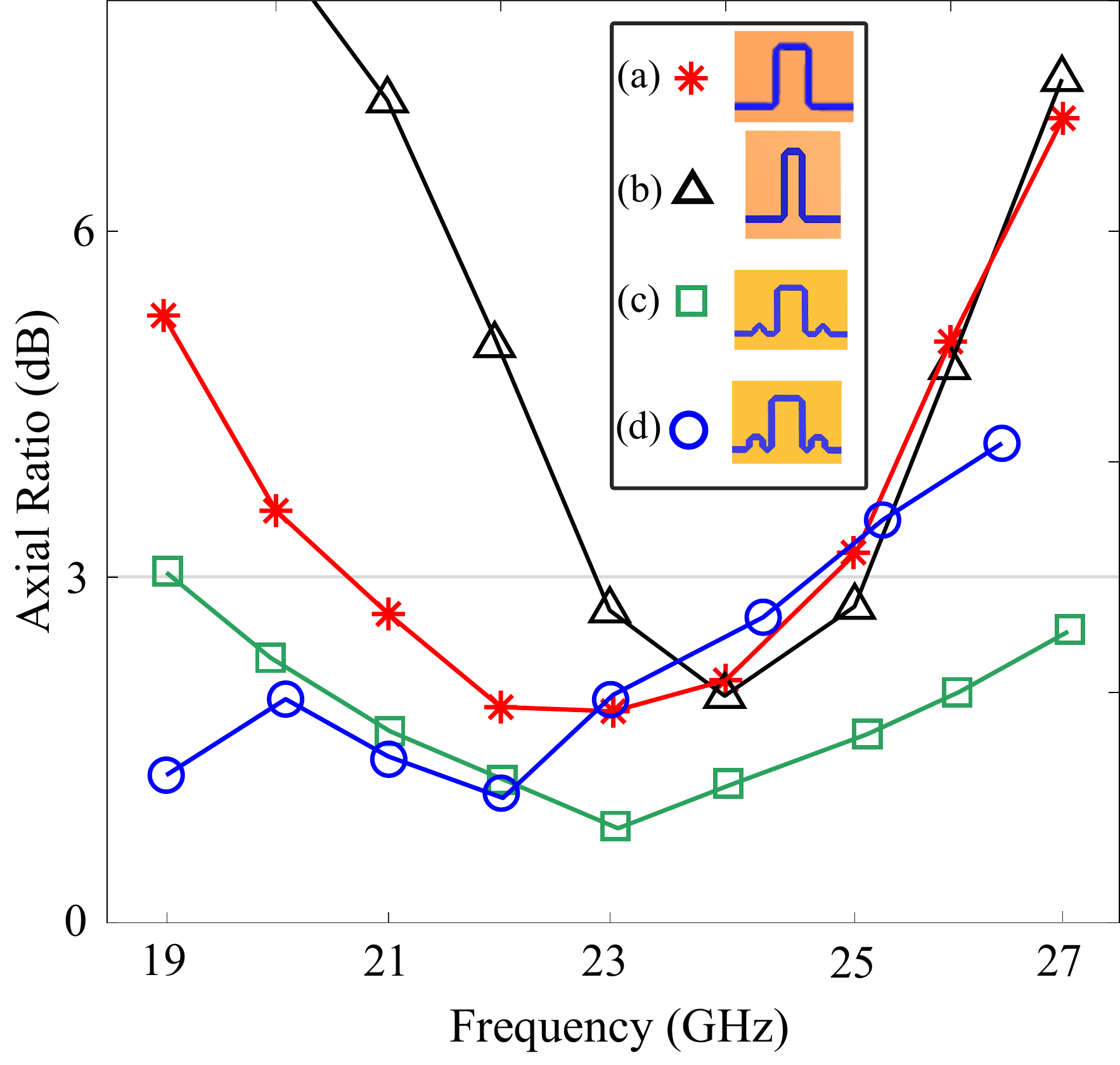}}
\caption{Axial ratio values in the main beam direction versus frequency obtained by full-wave simulation of 10 unit cells connected in series for 3 cases: (a)~single meander with $\phi_{\mathrm{AA_1}} = \pi$, (b)~single meander with $\phi_{\mathrm{AA_1}}=9\pi/8$ to reduce the period of the unit cell and {in presence of two additional meanders on either side of large meander (with $\phi_{\mathrm{AA_1}} = \pi$) with the size  (c)}~{($\phi_{\mathrm{sect}}= 10\degree$) and  (d)~($\phi_{\mathrm{sect}}= 25\degree$).}}
\label{TheoryandSimulation_CP}
\end{figure}

The corresponding axial ratio obtained through simulation at the angle of maximum gain is shown in Fig.~\ref{TheoryandSimulation_CP}(a). The antenna has good circular polarization performance near the design frequency of $f_\mathrm{b}=$ \SI{23}{\GHz}. The frequency range where the axial ratio is lower than \SI{3}{\dB} is from \SIrange{20}{25.2}{\GHz} with the beam scanning from ($-26\degree$ to $+10\degree$).

\mycomment{Hence there is some difference in axial ratio calculated theoretically and due to full-wave simulation. However, Qualitatively, the theoretical formulation gives a good idea of the circular polarization performance in operating range.}

\subsection{Improvement in the scanning range by reducing the period of the unit cell}
In the previous subsection, initial dimensions were taken such that $\phi_{\mathrm{AA_1}}=\pi$ and $\phi_{\mathrm{A_1B}}=\pi/2$ at $f=f_\mathrm{b}$. The scanning range and the separation of the harmonics can be increased by reducing the period of a unit cell. If the vertical length $l_\mathrm{AA_1}$ is increased such that $\phi_{\mathrm{AA_1}}=9\pi/8$, then according to \eqref{phidiffAB}, to maintain the circular polarization performance $\phi_{\mathrm{A_1B}}=\pi/4$ at the design frequency $f_\mathrm{b}$. This reduces the period of the unit cell since the horizontal interval ($l_\mathrm{A_1B}$) of the unit cell is shortened leading to better separation of harmonics. The rest of the dimensions remain the same as discussed in Table \ref{SingleUnitCellDimensions_Table}.

\begin{figure}[!t]
\centerline{\includegraphics[width=0.421\textwidth]{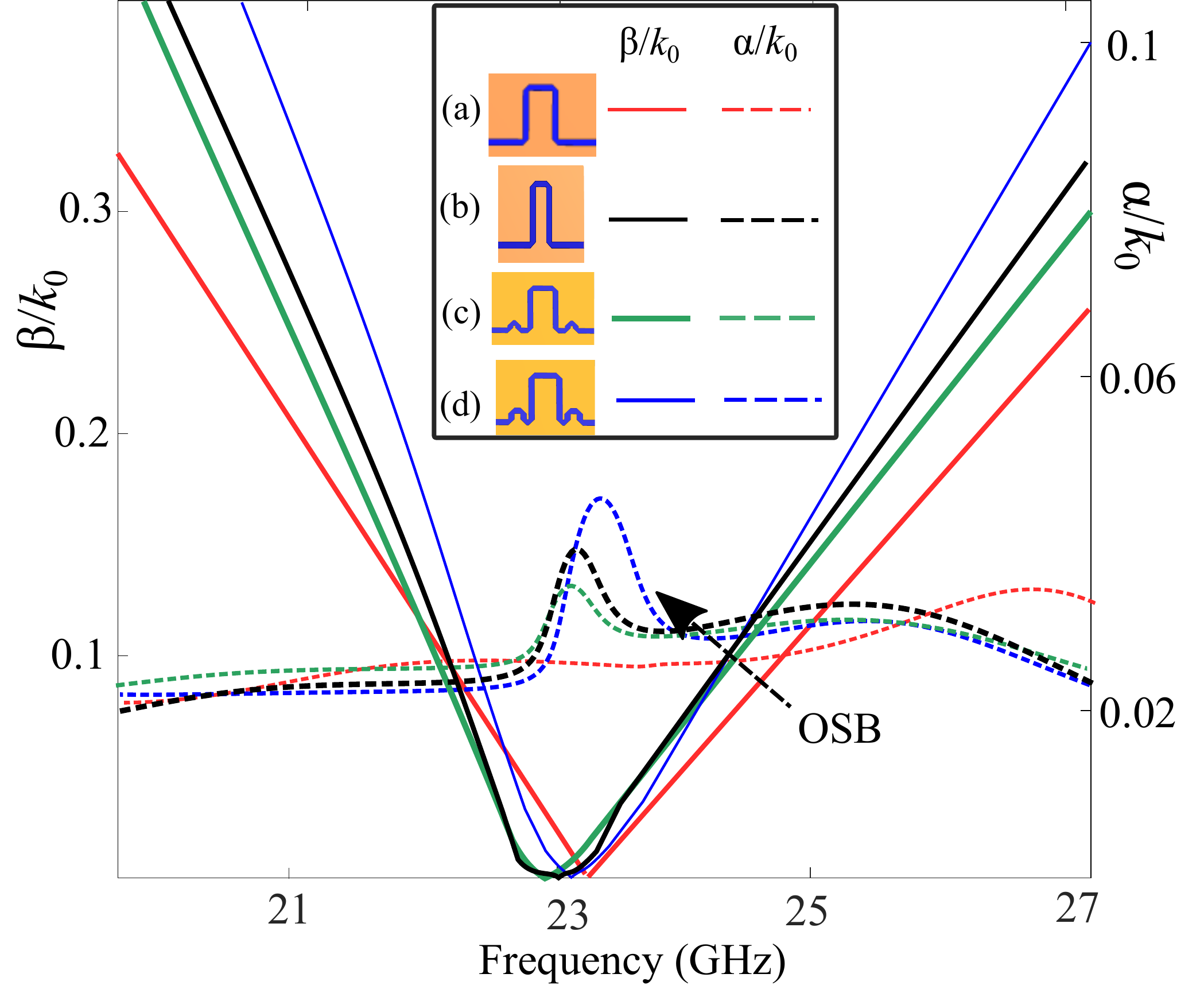}}
\caption{{Change in the dispersion parameters ($\beta$ and $\alpha$) for four cases: (a)~single meander with $\phi_{\mathrm{AA_1}} = \pi$. (b)~single meander with $\phi_{\mathrm{AA_1}}=9\pi/8$ to reduce the period of the unit cell, in the presence of two additional meanders on either side of large meander ($\phi_{\mathrm{AA_1}} = \pi$) with the size of (c)~($\phi_{\mathrm{sect}}~=~10\degree$) and (d)~($\phi_{\mathrm{sect}}~=~25\degree$). OSB expands with increase in size of additional meanders.}}
\label{ScanningRateandOSB}
\end{figure}

Ten such unit cells are connected in series to form a \gls{lwa} and full-wave simulations are performed in CST and Ansys HFSS. {This results in an increase in the scanning rate (the change in $\beta$ with change in frequency) as can be seen in Fig.~\ref{ScanningRateandOSB}(a, b).} However, as shown in Fig.~\ref{TheoryandSimulation_CP}(b), there is a negative impact on the circular polarization performance in the operating frequency band. 

In both the discussed cases with only a single meander in the unit cell [see Fig.~\ref{TheoryandSimulation_CP}(a, b)], at the design frequency $f=f_\mathrm{b}$ and in the frequency range close to it, the antenna shows good circular polarization performance. However, it deteriorates quite fast towards the lower and higher end of the band. Hence, the beam scanning range is pretty limited.

It can be concluded that using a unit cell with only a single large meander, is extremely challenging and may not be feasible to obtain large beam scanning range with good circular polarization performance.

\section{Improvement in the Scanning Range and Circular Polarization by Additional Meanders}
\label{sec:3meanders}
To enhance the beam steering range, improve circular polarization performance across the operational frequency band, and gain better control over scanning rate ($\Delta\theta/\Delta f$), two additional smaller meanders are introduced on either side of larger meander at an equal distance from the center of the unit cell. The following sections describe the characteristics and steps to design such a unit cell. 
 
\subsection{Constructing the unit cell with three meanders for dual-band operation}
Fig.\ref{E_abs_ver2}(e) depicts the design of improved unit cell. The first additional meander consists of three microstrip line intervals $\mathrm{CC_1}$, $\mathrm{C_1D}$ and $\mathrm{DD_1}$, while the third additional meander consists of line intervals $\mathrm{PP_1}$, $\mathrm{P_1Q}$ and $\mathrm{QQ_1}$. The lengths of each of these 6 microstrip line intervals is equal. The phase difference across each of these microstrip line intervals is defined to be $\phi_{\mathrm{sect}}$. Additionally, the phase difference due to the first and the third meander is denoted by $\phi_{\mathrm{TrSeg1}}$ and $\phi_{\mathrm{TrSeg3}}$, respectively.

The unit cell is designed to be symmetric, therefore the phase difference across the line intervals is $\phi_{\mathrm{D_1A}} = \phi_{\mathrm{B_1P}}$, $\phi_{\mathrm{OC}}=\phi_{\mathrm{Q_1M}}$, and $\phi_{\mathrm{TrSeg1}} = \phi_{\mathrm{TrSeg3}}$. Furthermore, the phase difference across the $1^{\mathrm{st}}$ meander and the $3^{\mathrm{rd}}$ meander can be written as 
\begin{subequations}
\begin{align}
        \phi_{\mathrm{TrSeg1}} &= \phi_{\mathrm{CC_1}} + \phi_{\mathrm{C_1D}}+\phi_{\mathrm{DD_1}} = 3\times\phi_{\mathrm{sect}}  \\
        \phi_{\mathrm{TrSeg3}} &= \phi_{\mathrm{PP_1}} + \phi_{\mathrm{P_1Q}}+\phi_{\mathrm{QQ_1}} = 3\times\phi_{\mathrm{sect}}
        \end{align}
\end{subequations}

In order to have broadside radiation at $f=f_\mathrm{b}$, a $4 \pi$ phase difference across the unit cell period $p$ must be satisfied. Consequently, from the Fig.~\ref{E_abs_ver2}(e), the design equation for the improved unit cell can be written as:
\begin{equation}
\begin{split}
    4\pi = &\phi_{\mathrm{OC}|f=f_\mathrm{b}}+\phi_{\mathrm{TrSeg1}|f=f_\mathrm{b}} + \phi_{\mathrm{D_1A}|f=f_\mathrm{b}} + \\ &\phi_{\mathrm{LargeSeg}|f=f_b} + \phi_{\mathrm{B_1P}|f=f_\mathrm{b}}+\\&\phi_{\mathrm{TrSeg3}|f=f_\mathrm{b}} + \phi_{\mathrm{Q_1M}|f=f_\mathrm{b}}  \label{threefingphase}
\end{split}
\end{equation}

Here, $\phi_{ij}$ represents the phase difference between the points $i$ and $j$ where $i, j \in \{\mathrm{C, C_1, D, D_1, A, A_1, B, B_1, P, P_1, Q, Q_1, O, M}\}$.

To maintain circular polarization, it is necessary that phase difference across larger meander $\phi_{\mathrm{LargeSeg}|f=f_\mathrm{b}}$ equals $5\pi/2$, as stated in equation \eqref{importantconstraint}. Therefore, the following design constraint can be derived from equation \eqref{threefingphase}:
\mycomment{
\begin{equation}
\begin{split}
    4\pi = \phi_{OC|f=f_b}+&\phi_{OA_1|f=f_b} +\phi_{AA_1|f=f_b}+\phi_{A_1B|f=f_b}+ \\ &\phi_{BB_1|f=f_b}+\phi_{B_1P|f=f_b} + \phi_{Q_1P|f=f_b}  
\end{split}
\end{equation}
}
\begin{equation}
   6\times\phi_{\mathrm{sect}|f=f_\mathrm{b}} +2\times\phi_{\mathrm{D_1A}|f=f_\mathrm{b}}+2\times\phi_{\mathrm{OC}|f=f_\mathrm{b}} = 3\pi/2 \label{constraint3fing1}
\end{equation}
\mycomment{INITIAL VALUES FROM "D:\Users\pvadher\Documents\CST Simulations\S5__TEMLeakyWave_NS\TEMLeakyWave_Model_Variation_3Fingers_WithLossMF_imp_CP_MF_ver8_LWA_ForPaper_no_t50.cst", FINAL VALUES: }
\mycomment{
\begin{table}
\begin{center}
\caption{Dimensions of the unit cell with 3 meanders designed to operate in spatial harmonic of $n=-2$ and have circularly polarized fields} \label{3FingUnitCellDimensions_Table}
\begin{tabular}[t]{p{0.30\linewidth}>{\centering}p{0.27\linewidth}>{\centering\arraybackslash}p{0.22\linewidth}>{\centering\arraybackslash}p{0.21\linewidth}}
\toprule
\textbf{Length Segment} & \textbf{In terms of $\lambda_b$} & \textbf{Initial Physical Values (mm)}& \textbf{Optimised Values (mm)} \\
\midrule
$OC=Q_1M$& $(\pi-3\phi_{sect})/2\times\lambda_b/(2\pi)$ & 1.819 & 0.376\\
$CC_1=C_1D=DD_1$& $\phi_{sect}\times\lambda_{b}/(2\pi)$ & 0.71 & 0.745\\
$D_1A=B_1P$ & $(\pi/2-3\phi_{sect})/2\times\lambda_{b}/(2\pi)$ & 1.128 & 1.172\\
$AA_1$ & $\lambda_b/2$ & 4.265 & 4.343\\
$A_1B$ & $\lambda_b/4$ & 2.383 & 3.062\\
$BB_1$ & $\lambda_b/2$ & 4.265 & 4.343\\
$PP_1=P_1Q=QQ_1$ & $\phi_{sect}\times\lambda_{b}/(2\pi)$ & 0.71 & 0.745\\
$p$ & $2\times\lambda_b$ & 9.694 & 8.028\\
$t_{50}$ &  & 0.5 & 0.5\\
$h_{sub}$ & & 0.254 & 0.254\\
\bottomrule
\end{tabular}
\end{center}
\end{table}
}

\captionsetup[table]{name=TABLE,labelsep=newline,textfont=sc, justification=centering}
\begin{table}[!t]
\renewcommand{\arraystretch}{1.3}
\centering
\caption{Dimensions of the unit cell with 3 meanders designed to operate in spatial harmonic of $n=-2$ and have circularly polarized fields. Note that $\lambda_\mathrm{b}$ is wavelength corresponding to $f_\mathrm{b}$ in microstrip medium} 
\label{3FingUnitCellDimensions_Table}
\begin{tabular}{p{0.22\linewidth}>{\centering}p{0.37\linewidth}>{\centering\arraybackslash}p{0.08\linewidth}>{\centering\arraybackslash}p{0.11\linewidth}}
\toprule
    \multirow{3}*{\thead*{Line\\interval}}
    & \multicolumn{2}{c}{\thead{Initial Values (From the theory)}}
    & \multirow{3}*{\thead{Optimised\\(mm)}} \\
\cmidrule(lr){2-3}
     & \thead{Expressed in $\lambda_\mathrm{b}$} & \thead{(mm)} &  \\
\midrule
$\mathrm{CC_1}$=$\mathrm{C_1D}$=$\mathrm{DD_1}$& $\phi_{\mathrm{sect}}\times(\lambda_{\mathrm{b}}/2\pi)$ & 0.71 & 0.745\\
$\mathrm{D_1A}$=$\mathrm{B_1P}$ & $(\pi/2-3\phi_{\mathrm{sect}})/2\times(\lambda_{\mathrm{b}}/2\pi)$ & 1.128 & 1.172\\
$\mathrm{AA_1}$ & $\lambda_\mathrm{b}/2$ & 4.265 & 4.343\\
$\mathrm{A_1B}$ & $\lambda_\mathrm{b}/4$ & 2.383 & 3.062\\
$\mathrm{BB_1}$ & $\lambda_\mathrm{b}/2$ & 4.265 & 4.343\\
$\mathrm{PP_1}$=$\mathrm{P_1Q}$=$\mathrm{QQ_1}$ & $\phi_{\mathrm{sect}}\times(\lambda_{\mathrm{b}}/2\pi)$ & 0.71 & 0.745\\
$\mathrm{OC}$=$\mathrm{Q_1M}$& $(\pi-3\phi_{\mathrm{sect}})/2\times(\lambda_\mathrm{b}/2\pi)$ & 1.819 & 0.376\\
$p$ &$2\times\lambda_\mathrm{b}$ & 9.694 & 8.028\\
$\mathrm{t}_{50}$ & - & 0.5 & 0.5\\
$\mathrm{h}_{\mathrm{sub}}$ & - & 0.254 & 0.254\\
\bottomrule
\end{tabular}
\end{table}

\subsection{Circular polarization analysis for the improved unit cell}\label{subsec:CP_improved}
It was proved in Section \ref{sec:radiationmechanism} that a single meander unit with four radiating elements can produce circular polarization at the desired design frequency. Hence, for the improved unit cell with 12 radiating currents (see Fig.~\ref{E_abs_ver2}(e)), circular polarization at $f = f_{\mathrm{b}}$ can be achieved if the effect due to radiating currents on the first additional meander (at $\mathrm{C}$, $\mathrm{C_1}$, $\mathrm{D}$ and $\mathrm{D_1}$) and the third additional meander (at $\mathrm{P}$, $\mathrm{P_1}$, $\mathrm{Q}$ and $\mathrm{Q_1}$) is reduced. This is possible if the direction of current at $\mathrm{C}$ is opposite to the current at $\mathrm{P}$ resulting in opposing fields at these corners.

\mycomment{The radiation due to the mitred corners is the result of magnetic current sources ($M$). Hence from \cite{harrington_introduction_1961}, 
\begin{subequations}
\begin{align}
    M_\mathrm{Y'| C} &= E_{\mathrm{X'|C}}\times \hat{z}' \\
    M_\mathrm{Y'| P} &= E_{\mathrm{X'|P}}\times \hat{z}'    
\end{align}
\end{subequations}
where $\hat{z}'$ is normal to the plane of periodic antenna. To simplify the far field expressions, periodic antenna is considered with infinite unit cells is considered.}

The magnetic currents at the corners $\mathrm{C}$ and $\mathrm{P}$ are oriented along the ${\mathrm{Y'}}$ axis as shown in Fig.~\ref{thetavsfreq}(b), but depending on the phase shift introduced by the interconnecting microstrip lines, they can have the same or opposite directions. Hence, the length of the microstrip line intervals in between these two corners can be optimized in such a way that the magnetic currents are in opposite direction at design frequency $f=f_\mathrm{b}$. 

\mycomment{To simplify the far field expressions, periodic antenna is considered with infinite unit cells sequentially arranged in the Y-direction (see Fig.~\ref{thetavsfreq}(b)).} 

\mycomment{The phases at $\mathrm{C}$ and $\mathrm{P}$ with respect to starting point \ma, respectively are:
\begin{subequations}
\begin{align}
    \phi_\mathrm{C} &= \phi_{\mathrm{OC}} \\
    \label{}
    \phi_\mathrm{P} &= \phi_{\mathrm{OC}}+\phi_{\mathrm{TrSeg1}}+\phi_{\mathrm{D_1A}}+\phi_{\mathrm{LargeSeg}}+\phi_{\mathrm{B_1P}}
    \label{}
    \end{align}
\end{subequations}}

\mycomment{In far field, in the system origin the ratio of the magnetic currents due to the corners at C and P can be written as 
\begin{equation}
    \frac{M_{\mathrm{X'|C}}}{M_{\mathrm{X'|P}}} = \frac{\mathrm{e}^{-j\phi_\mathrm{C}}}{\mathrm{e}^{-j\phi_\mathrm{P}}} = \mathrm{e}^{-j(\phi_\mathrm{C}-\phi_\mathrm{P})}
\end{equation}}

The direction of magnetic current at $\mathrm{C}$ will be opposite to the magnetic current at $\mathrm{P}$ if the following constraint is met at $f=f_\mathrm{b}$ (since $\mathrm{e}^{3\pi}=-1$): 
\begin{subequations}
\begin{align}
    \phi_{\mathrm{OP}}-\phi_{\mathrm{OC}} &= 3\pi \\
    \phi_{\mathrm{TrSeg1}}+\phi_{\mathrm{D_1A}}+\phi_{\mathrm{LargeSeg}}+\phi_{\mathrm{B_1P}} &= 3\pi
\end{align}
\end{subequations}
\mycomment{ 
\begin{subequations}
\begin{align}
    &\phi_{\mathrm{P}|f=f_\mathrm{b}}-\phi_{\mathrm{C}|f=f_\mathrm{b}} &= 3\pi \\
        &\phi_{\mathrm{TrSeg1}|f=f_\mathrm{b}}+\phi_{\mathrm{D_1A}|f=f_\mathrm{b}}+\phi_{\mathrm{LargeSeg}|f=f_\mathrm{b}}+\phi_{\mathrm{B_1P}|f=f_\mathrm{b}} &= 3\pi
\end{align}
\end{subequations}
}

From equation \eqref{importantconstraint}, it is known that to maintain circular polarization at $f=f_\mathrm{b}$, $\phi_{\mathrm{LargeSeg}} =5\pi/2$, hence the following design equation is obtained
\begin{equation}
    3\times\phi_{\mathrm{sect}|f=f_\mathrm{b}} + 2\times\phi_{\mathrm{D_1A}|f=f_\mathrm{b}} = \pi/2
    \label{constraint3fing2}
\end{equation}

The equations \eqref{importantconstraint}, \eqref{constraint3fing1} and \eqref{constraint3fing2} are the design constraints on the length of line intervals for the proposed unit cell to maintain broadside radiation and circular polarization at $f~=~f_\mathrm{b}$.

From equation \eqref{importantconstraint}, it is clear that the dimensions for larger meander are fixed to maintain circular polarization. The other dimensions are governed by equations \eqref{constraint3fing1} and \eqref{constraint3fing2}. This results in 3 variables (namely $\phi_{\mathrm{sect}|f=f_\mathrm{b}}$, $\phi_{\mathrm{D_1A}|f=f_\mathrm{b}}$ and $\phi_{\mathrm{OC}|f=f_\mathrm{b}}$) and 2 equations. In the current analysis $\phi_{\mathrm{sect}|f=f_\mathrm{b}}$ is considered an independent variable making the other two variables ($\phi_{\mathrm{D_1A}|f=f_\mathrm{b}}$ and $\phi_{\mathrm{OC}|f=f_\mathrm{b}}$) dependent. Initial design is chosen such that $\phi_{\mathrm{sect}|f=f_\mathrm{b}} = 10^{\degree}$. The other two dimensions can hence be calculated from equation \eqref{constraint3fing1} and equation \eqref{constraint3fing2}. The initial lengths of the microstrip sections following this theory are shown in the table \ref{3FingUnitCellDimensions_Table} for $f_\mathrm{b}=$ \SI{23}{\GHz}. Furthermore, the placements of small meanders with respect to large meander within the unit cell can be varied as well to obtain better circular polarization.

\begin{figure}[!t]
\centerline{\includegraphics[width=0.4\textwidth]{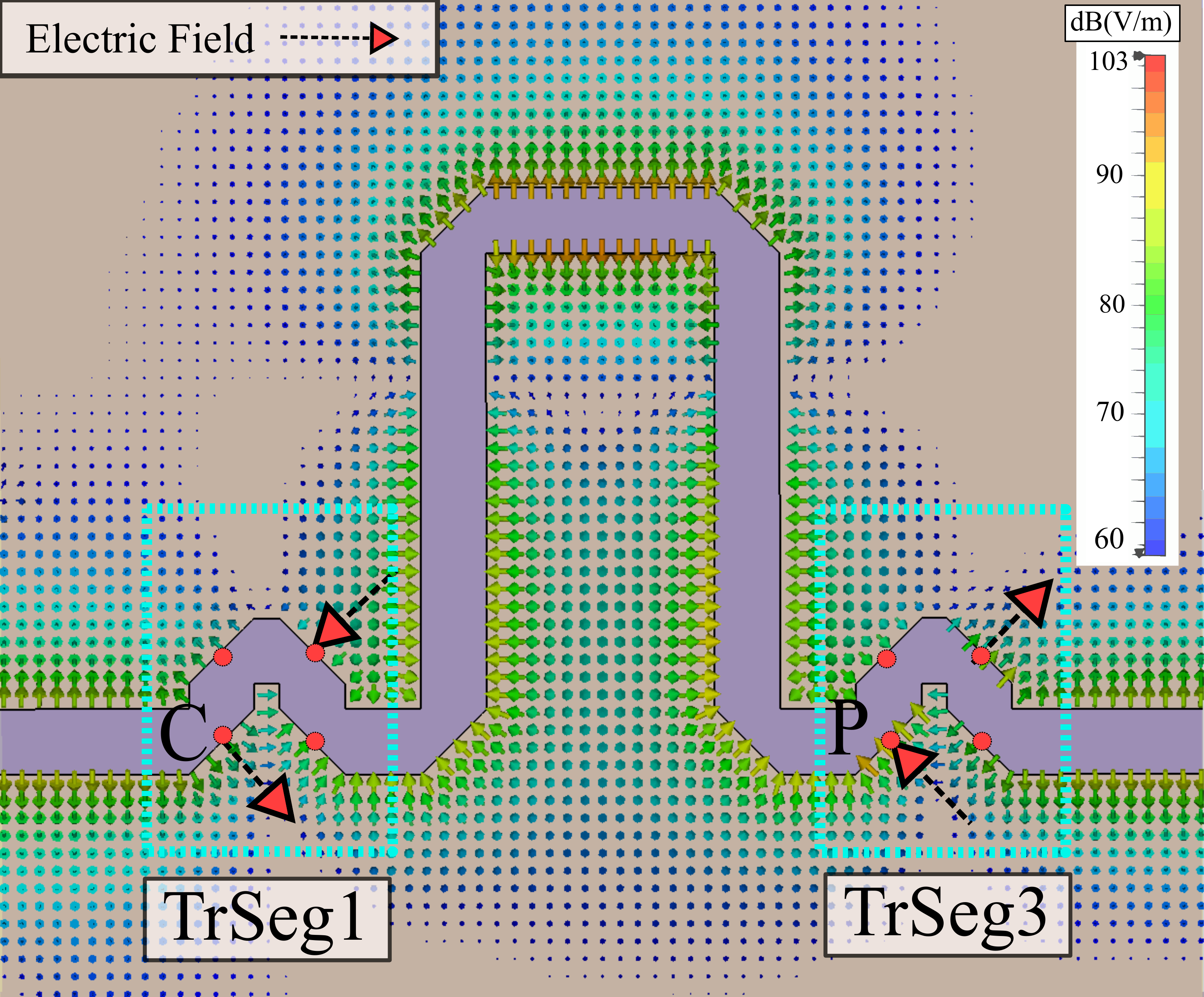}}
\caption{Full-wave simulation for the unit cell constructed using the theoretical formulation. At $f=f_\mathrm{b}$, the electric fields at each of the four mitred corners of $\mathrm{TrSeg1}$ cancels out with corresponding mitred corner of $\mathrm{TrSeg3}$.}
\label{unitcellcurrent}
\end{figure}

 Fig.~\ref{unitcellcurrent} shows the full-wave simulation of the unit cell at $f=f_\mathrm{b}$ designed according to the equations above. As can be seen, the direction of fields at each of the corners in the $1^{\mathrm{st}}$ smaller meander is opposite to the field at the corresponding corner of the $3^{\mathrm{rd}}$ smaller meander.

\mycomment{The optimised design values following the analysis from simulation and theory described above is also listed.}

10 such unit cells are sequentially connected as shown in Fig.~\ref{thetavsfreq}(b) to form a \gls{lwa} and then simulated in CST. Fig.~\ref{TheoryandSimulation_CP}(c) shows the unit cell (in the inset) designed by the formulated theory described above and simulation respectively, of axial ratio obtained for different frequencies over the scanning range for the main fan beam direction emanated from the \gls{lwa}. Notable improvement in axial ratio performance over a larger frequency band can be observed following the inclusion of two smaller meanders.
{
\subsection{Improvement in scanning range and better control of~scanning rate}\label{subsec:scanningrate}
The size of the two additional meanders can be changed to modify the period ($p$) of the unit cell. Fig. \ref{ScanningRateandOSB}(a, c, d) shows the impact of the introduction and size of additional meanders on the dispersion parameters $\alpha$ and $\beta_{-2}$ (radiating harmonic). As can be observed, the slope of $\beta_{-2}$ as a function of frequency (and hence the beam scanning range of the \gls{lwa}) can be controlled efficiently by introducing and varying the size of additional meanders.}

{In Fig.~\ref{TheoryandSimulation_CP}(a, c, d), the impact on circular polarization due to alterations in meander size can be observed across a wide range of frequencies in K-band. The circular polarization at the design frequency $f_\mathrm{b}$ remains intact. Increasing the meander size increases the scanning rate ($\Delta\beta_{-2}/\Delta f$), however the circular polarization over the large band of operation (K-band) is impacted negatively. Hence in the fabricated prototype the size of small meander corresponding to $\phi_\mathrm{sect}=10^\degree$ is chosen.}

\begin{figure*}[!t]
\centerline{\includegraphics[width=0.98\textwidth]{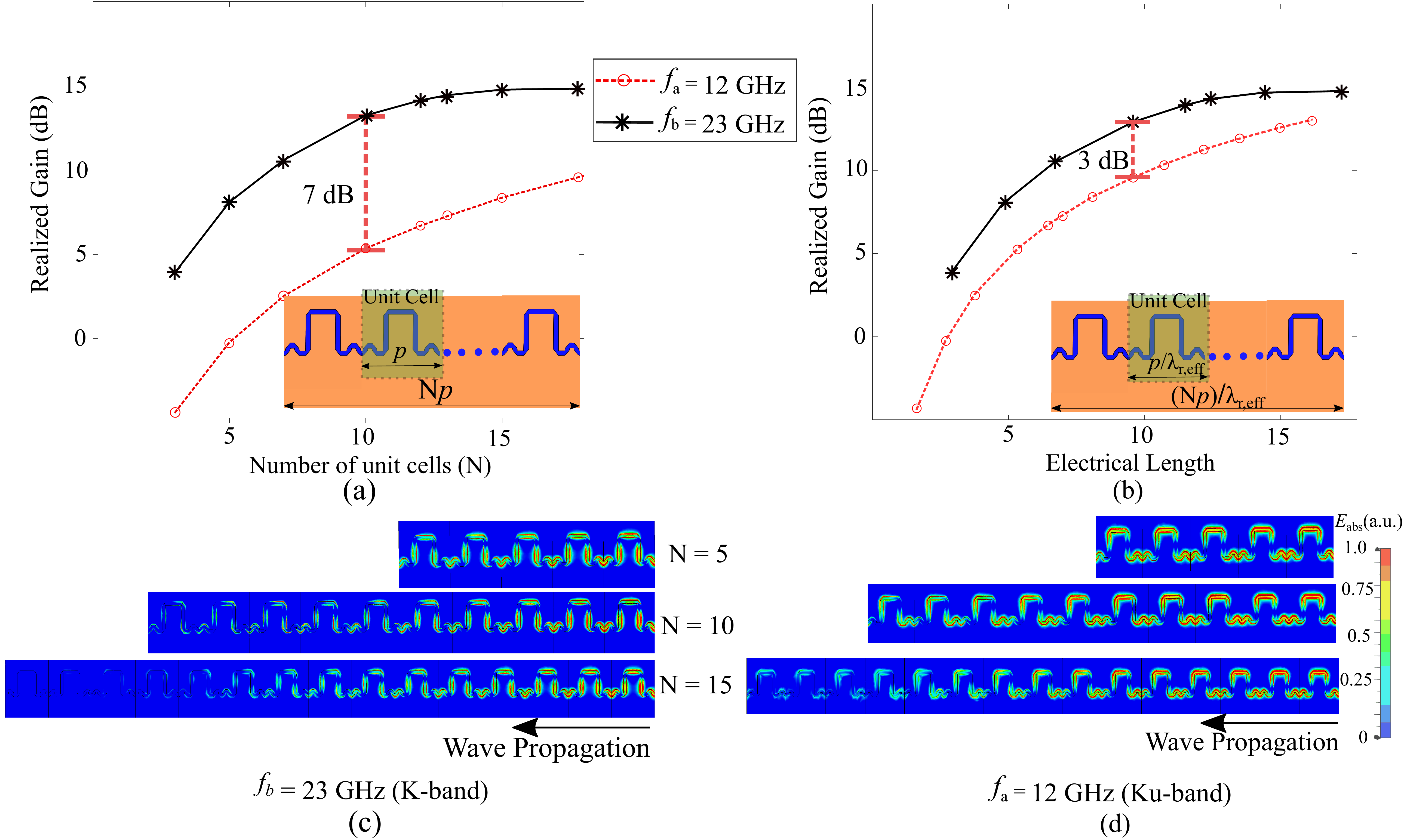}}
\caption{{Increase in the maximum realized gain at $f_\mathrm{a}=$\SI{12}{GHz} and $f_\mathrm{b}=$\SI{23}{GHz} with the increase in (a) physical length and (b)~electrical length. Magnitude of electric field along the LWA at (c)~$f=f_\mathrm{b}$, (d) $f=f_\mathrm{a}$.}}
\label{RGvsEL}
\end{figure*}

\subsection{{Overall p}erformance improvement due to the two additional meanders}
Due to the two additional smaller meanders the electrical length of the unit cell has been increased. This results in following improvements for frequency scanning:
\begin{itemize}
    \item enhanced scanning range,
    \item better control over scanning rate by dictating the size of path--length in the unit cell by reducing or increasing the size of small meanders ($\phi_{\mathrm{sect}}$),
    \item better separation of spatial harmonics resulting in lower spurious radiation due to non-desired spatial harmonics, 
    \item improvement of circular polarization performance across the band.
\end{itemize}

By comparing the Brillouin diagrams in Fig.~\ref{E_abs_ver2}(c) and Fig.~\ref{E_abs_ver2}(e) for the unit cell geometry containing one meander and three meanders, respectively, the first three conclusions can be easily drawn. With the help of additional two meanders the rate of dispersion with frequency is controlled within the unit cell. The impact of the proposed geometry on the circular polarization can be observed from Fig.~\ref{TheoryandSimulation_CP}. The acceptable range for frequency scanning (frequency range where axial ratio $<$~\SI{3}{\dB}) has increased from \SIrange{20}{25.2}{\GHz} to \SIrange{19.4}{27.5}{\GHz} resulting in increase of beam steering from (\SI{-26}{\degree} to \SI{+10}{\degree}) to (\SI{-42}{\degree} to \SI{+ 30}{\degree}). The beam steering (\SI{-42}{\degree} to \SI{+ 30}{\degree}) is discussed in Section \ref{sec:measurement}.
{\subsection{Determining the number of unit cells for LWA}\label{subsec:numofunitcells}}
{To keep the antenna design compact, determining the length of the \gls{lwa} $(=\mathrm{N}p)$ such that it radiates most of the power for the least number of unit cells ($\mathrm{N}$) is quite important. Fig.~\ref{RGvsEL}(a) shows the change in gain as a function of the number of unit cells at the design frequency $f_\mathrm{b}=$\SI{23}{\GHz}. The rate of gain change decreases as the number of unit cells exceeds 10. The plot of electric field magnitude along the antenna [Fig.\ref{RGvsEL}(c)] shows that the antenna radiates most of the power when $\mathrm{N}=$10.}

{Likewise, the magnitude of electric field at $f_\mathrm{a}=$ \SI{12}{GHz} (broadside frequency in the Ku-band) is shown in Fig.~\ref{RGvsEL}(d). For $f_\mathrm{a}$, there is high residual power at the output port when only 10 unit cells are cascaded to form \gls{lwa}. The antenna is electrically smaller in Ku-band than in K-band, thus leading to a higher gain in K-band. The difference in gain is close to \SI{7}{dB} [Fig.~\ref{RGvsEL}(a)].}

{
However, as shown in Fig.~\ref{RGvsEL}(b), the gain difference is less than \SI{3}{\dB} between the Ku-band and K-band of operation when equal electrical length is considered. The electrical length of the \gls{lwa} is equal to $\mathrm{N}p/\lambda_\mathrm{r, eff}$, where $\lambda_\mathrm{r, eff}$ is the corresponding wavelength in the microstrip medium at $f_\mathrm{a}$ and $f_\mathrm{b}$ respectively. Since the antenna in this work is optimised to operate in the K-band, only 10 unit cells are cascaded in the designed prototype.}
\bigskip
\mycomment{\begin{figure}[h]
\centerline{\includegraphics[width=0.45\textwidth]{Figures/RGvsN.pdf}}
\caption{Change in the maximum realized gain at $f_\mathrm{a}=$ \SI{12}{GHz} and $f_\mathrm{b}=$\SI{23}{GHz} in terms of physical length.}
\label{RGvsN}
\end{figure}

\begin{figure}[h]
\centerline{\includegraphics[width=0.45\textwidth]{Figures/RGvsEL.pdf}}
\caption{Change in the maximum realized gain at $f_\mathrm{a}~=~$ \SI{12}{GHz} and $f_\mathrm{b}~=~$ \SI{23}{GHz} in terms of electrical length.}
\label{RGvsEL}
\end{figure}
}

\subsection{Mitigation of OSB}
Periodic structures like \glspl{lwa} have a frequency band near the broadside frequency ($f_\mathrm{b}$ in current case), where there is no conduction of the travelling-wave, known as \gls{osb} \cite{volakis_antenna_2007, balanis_antenna_2015}. To remove \gls{osb}, the parameter of Bloch impedance ($Z_{\mathrm{s}}$) has to be analysed. The Bloch impedance can be calculated from the S-parameters extracted from driven-mode full-wave simulation in terms of circuit parameters $\mathrm{A}$, $\mathrm{B}$, $\mathrm{C}$ and $\mathrm{D}$ as shown in \cite{pozar_microwave_2011}:
\begin{equation}
Z_{\mathrm{s}}= \frac{-2\times\mathrm{B}}{(\mathrm{A}-\mathrm{D}-\sqrt{(\mathrm{A}+\mathrm{D})^2-4)}} \label{zbloch}
\end{equation}

\begin{figure}[!t]
\centerline{\includegraphics[width=0.39\textwidth]{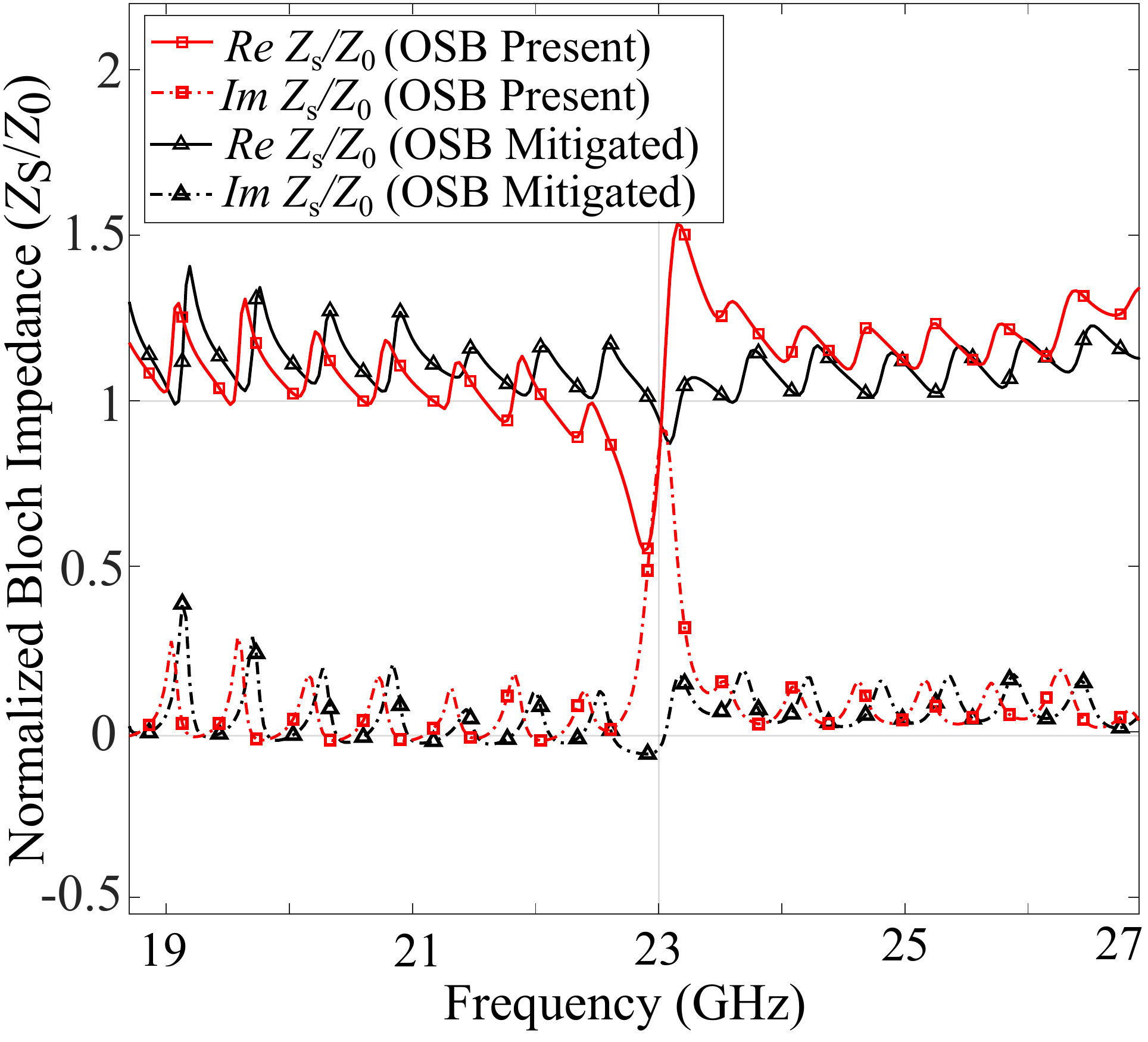}}
\caption{The Bloch impedance of the LWA when OSB is present and when OSB is mitigated by the introduction of additional capacitance. Here the impedance is normalised to $Z_0$ = $50 \Omega$.}
\label{Z_bloch_OSBRemoval}
\end{figure}

Fig.~\ref{Z_bloch_OSBRemoval} shows the Bloch impedance extracted for the K-band operation range. For the case when \gls{osb} is present, at the broadside frequency, $f_\mathrm{b}=\SI{23}{\giga\hertz}$, there is an abrupt increase in imaginary $Z_{\mathrm{s}}$, which results in poor transmission and high return loss. This results in reduced gain at the broadside region. Furthermore, while \gls{osb} is present at $f_\mathrm{a}=\SI{12}{\giga\hertz}$ and at $f_\mathrm{b}=\SI{23}{\giga\hertz}$, its impact is more pronounced at $f_\mathrm{b}$. Hence more emphasis is placed in the \gls{osb} suppression in the K-band.

\begin{figure}[!t]
\centerline{\includegraphics[width=0.35\textwidth]{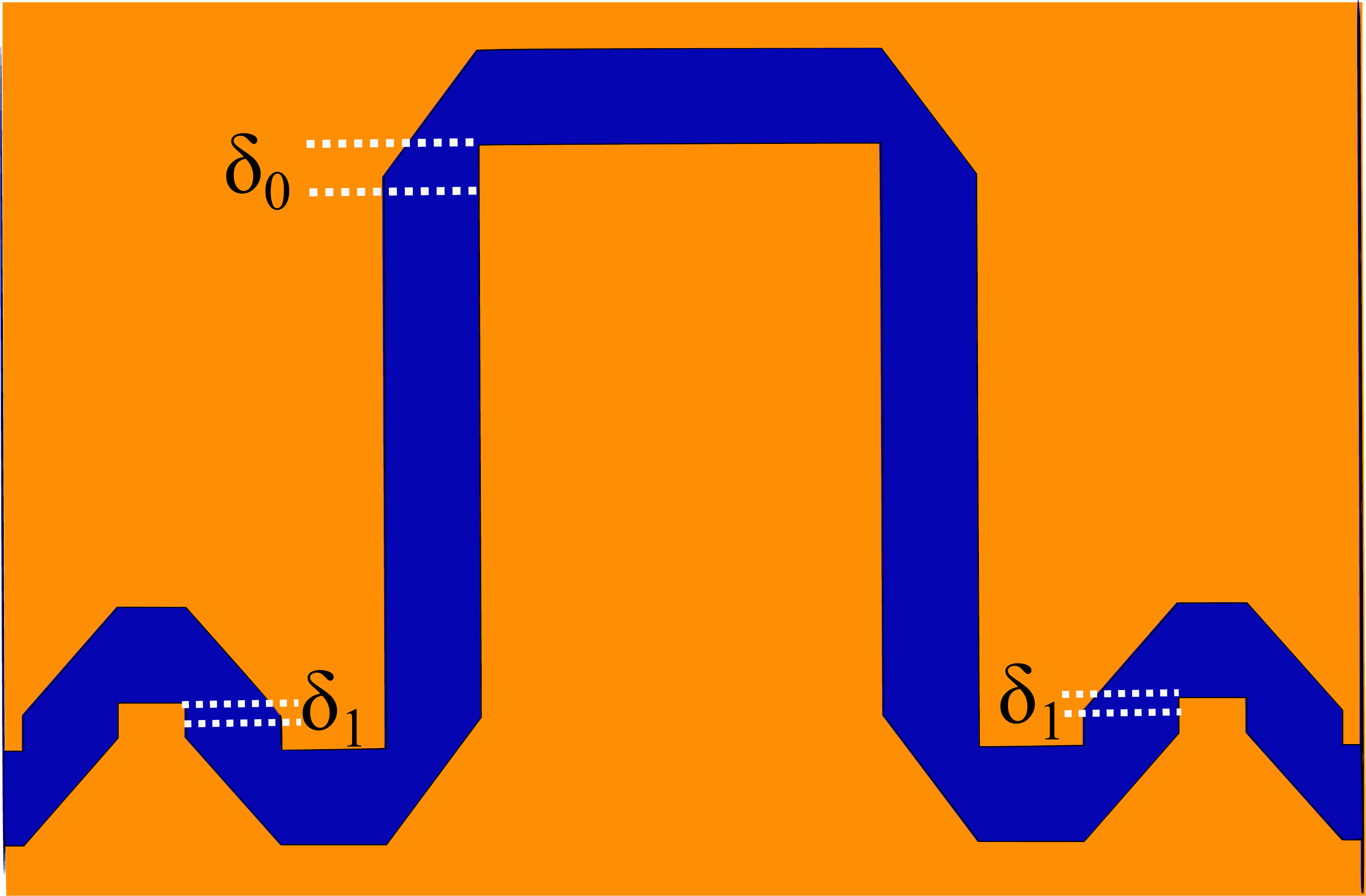}}
\caption{OSB mitigation by changing the angle of mitred corner. The values of $\delta_0$ and $\delta_1$ are \SI{0.15}{mm} and \SI{0.06}{mm} respectively.}
\label{UnitCellDistributionDimensions_OSB}
\end{figure}

To mitigate the problem of \gls{osb}, a technique similar to \cite{wang_periodic_2022} is employed. For a meandering microstrip with mitred corners the effect of bending the microstrip line (corners) can be modelled as capacitance \cite{douville_experimental_1978, silvester_microstrip_1973, anders_microstrip_1980} . This is clearly evident from Fig.~\ref{Z_bloch_OSBRemoval} when \gls{osb} is present. Hence, the angle for the mitred corners are changed which introduces an additional inductance \cite{anders_microstrip_1980, silvester_microstrip_1973} that reduces imaginary impedance at $f=f_\mathrm{b}$ as shown in Fig.~\ref{UnitCellDistributionDimensions_OSB}. Fig.~\ref{Z_bloch_OSBRemoval} shows that imaginary impedance goes to zero which leads to mitigation of \gls{osb} at broadside direction. 

\mycomment{Here one thing to note is that the variation of the mitered corners will create an impact the circular polarization performance. however, the mitred corners for individual segment are at equal angle and similar analysis has been carried out as described in Section \ref{subsec:CP_improved} theoretically and through simulations. }


\begin{figure}[!t]
\centerline{\includegraphics[width=0.4\textwidth]{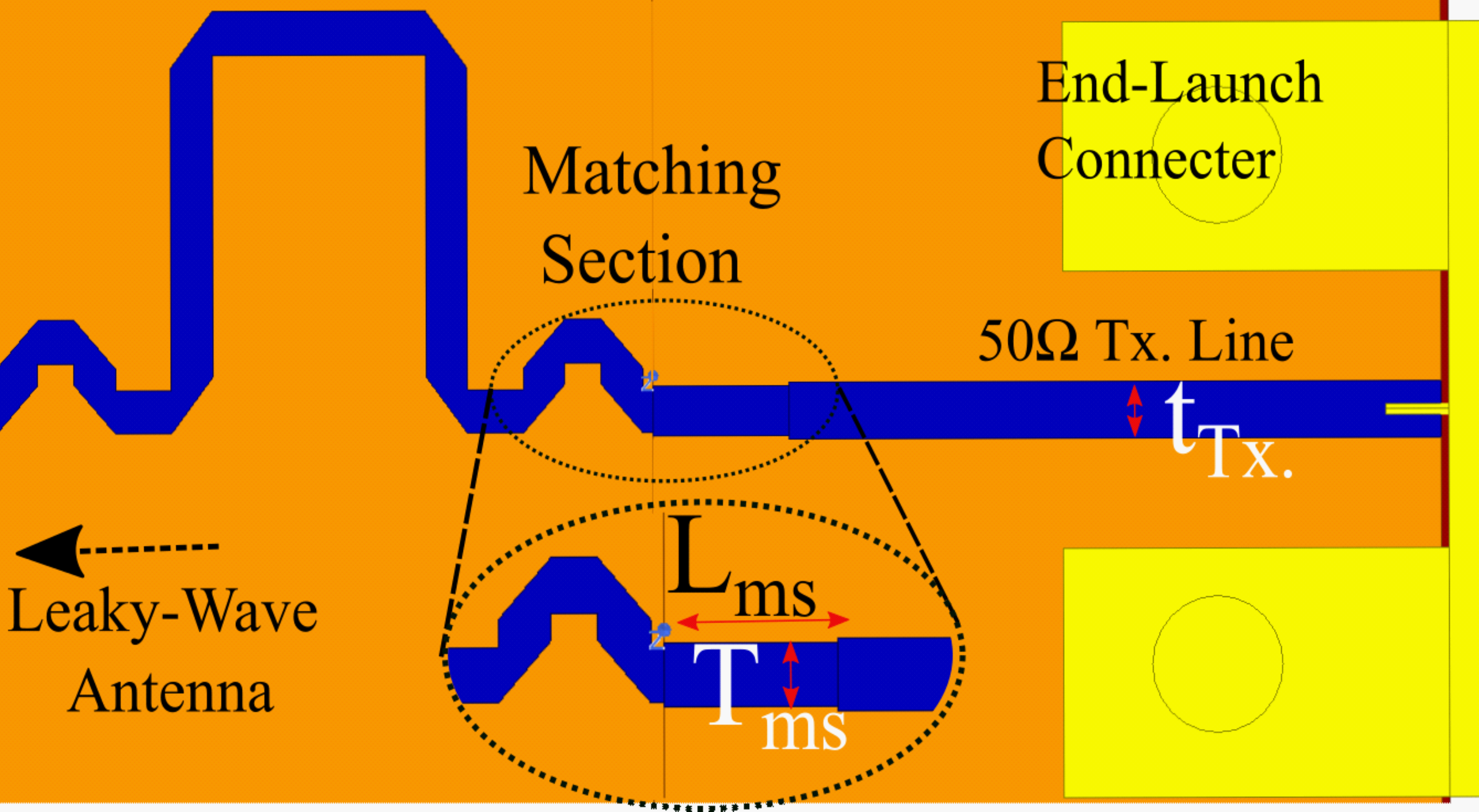}}
\caption{Feed Design to match the output impedance ($Z_0$) of $50 \Omega$ connector to input impedance of of antenna at around $60\Omega$. The values for matching section is $\mathrm{T_{ms}}=~\SI{0.57}{\milli\meter}$ and $\mathrm{L_{ms}}=~\SI{1.634}{\milli\meter}$. The $50 \Omega$ transmission line is $\mathrm{t_{Tx}}=~\SI{0.64}{\milli\meter}$.}
\label{FeedStructure}
\end{figure}

\begin{figure}[!t]
\centerline{\includegraphics[width=0.5\textwidth]{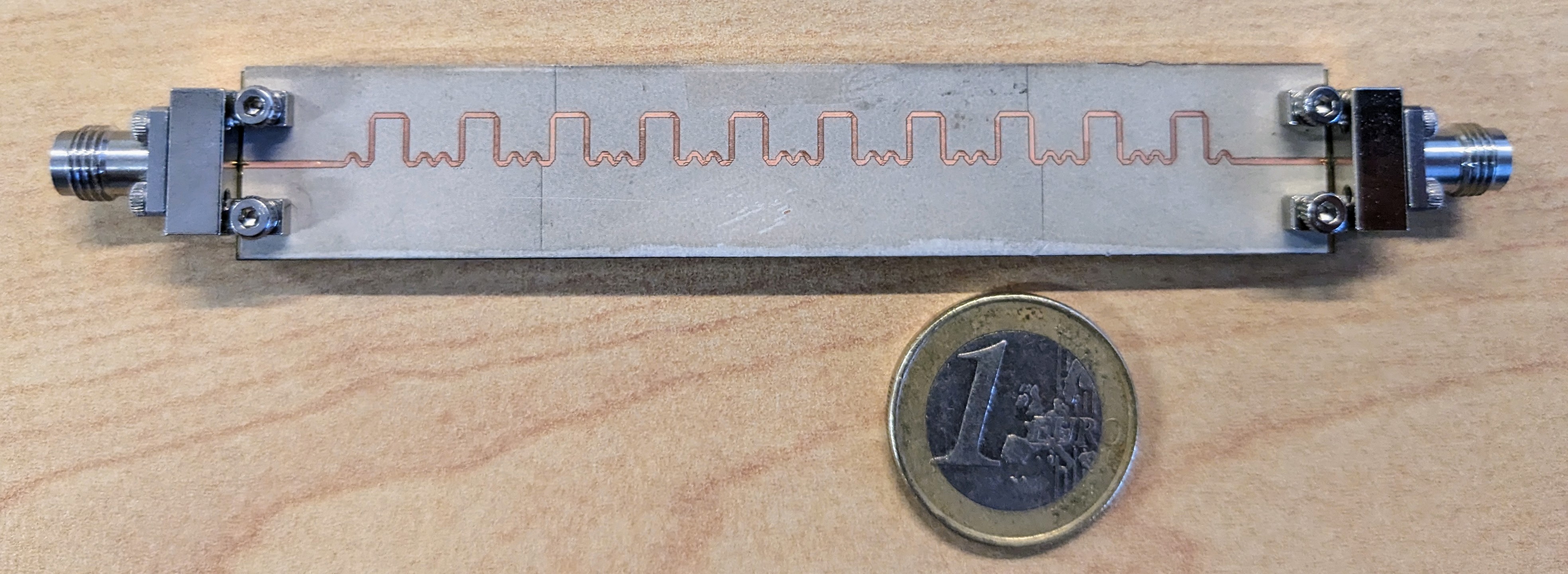}}
\caption{Fabricated Leaky-Wave Antenna.}
\label{FabricatedAntenna}
\end{figure}

\begin{figure*}[!h]
\centerline{\includegraphics[width=1\textwidth]{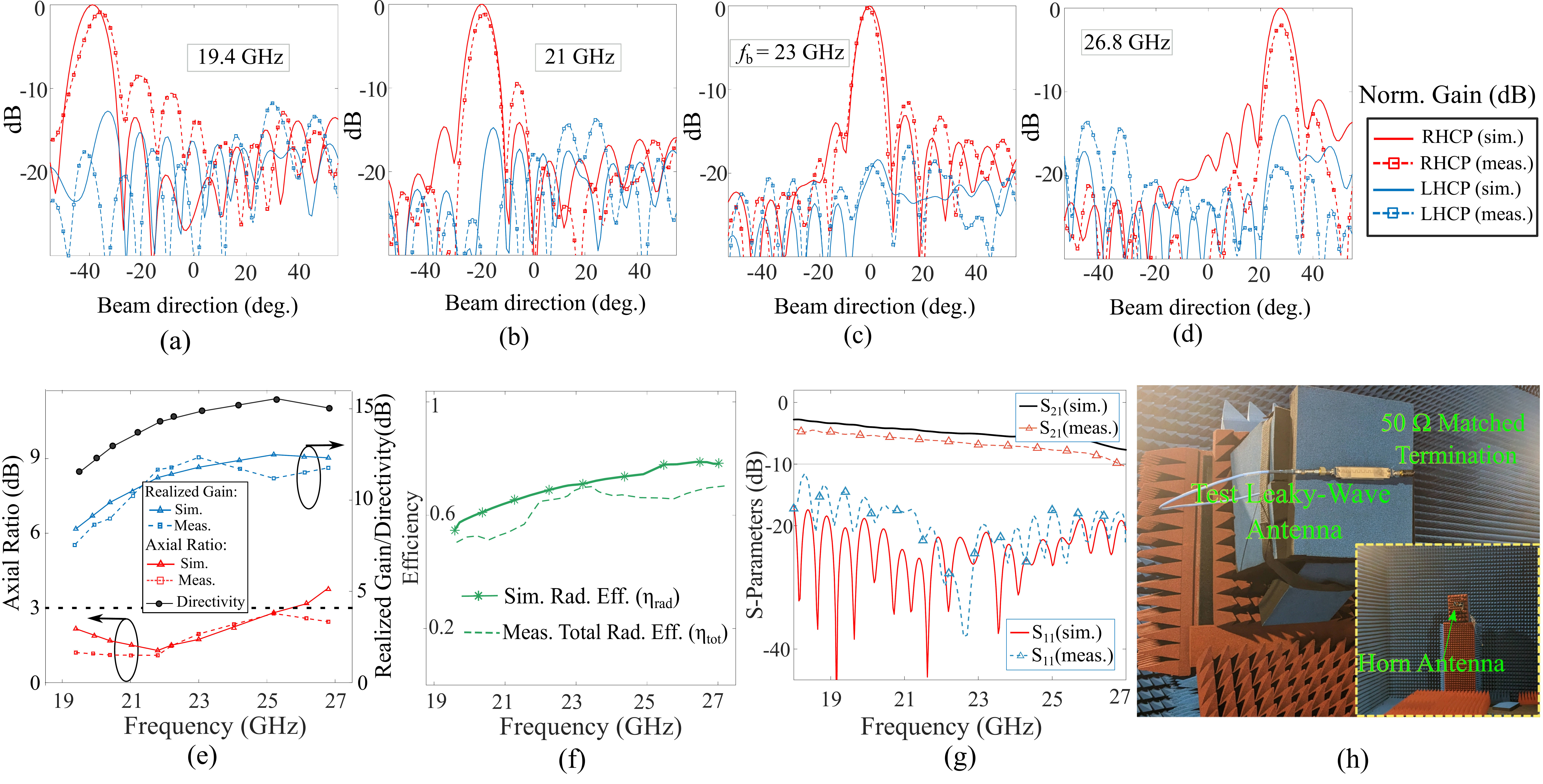}}
\caption{Measurements in K-band where the antenna is circularly polarized. Measured and simulated normalized radiation pattern of the fabricated antenna in the H-plane at (a) \SI{19.4}{GHz}, (b) \SI{21}{GHz}, (c) \SI{23}{GHz}(${=f_\textrm{b}}$), and (d) \SI{26.8}{GHz}. The antenna scans from \SIrange{-42}{30}{\degree} in the K-band frequency range. (e)~Axial ratio, realized gain and directivity of the fabricated antenna.~{(f)~Efficiency of the designed antenna} (g)~{Measured and simulated S-parameters}. (h)~Measurement setup of the antenna.\\}
\label{KBandMeasure}
\end{figure*}

\begin{figure*}[!h]
\centerline{\includegraphics[width=0.98\textwidth]{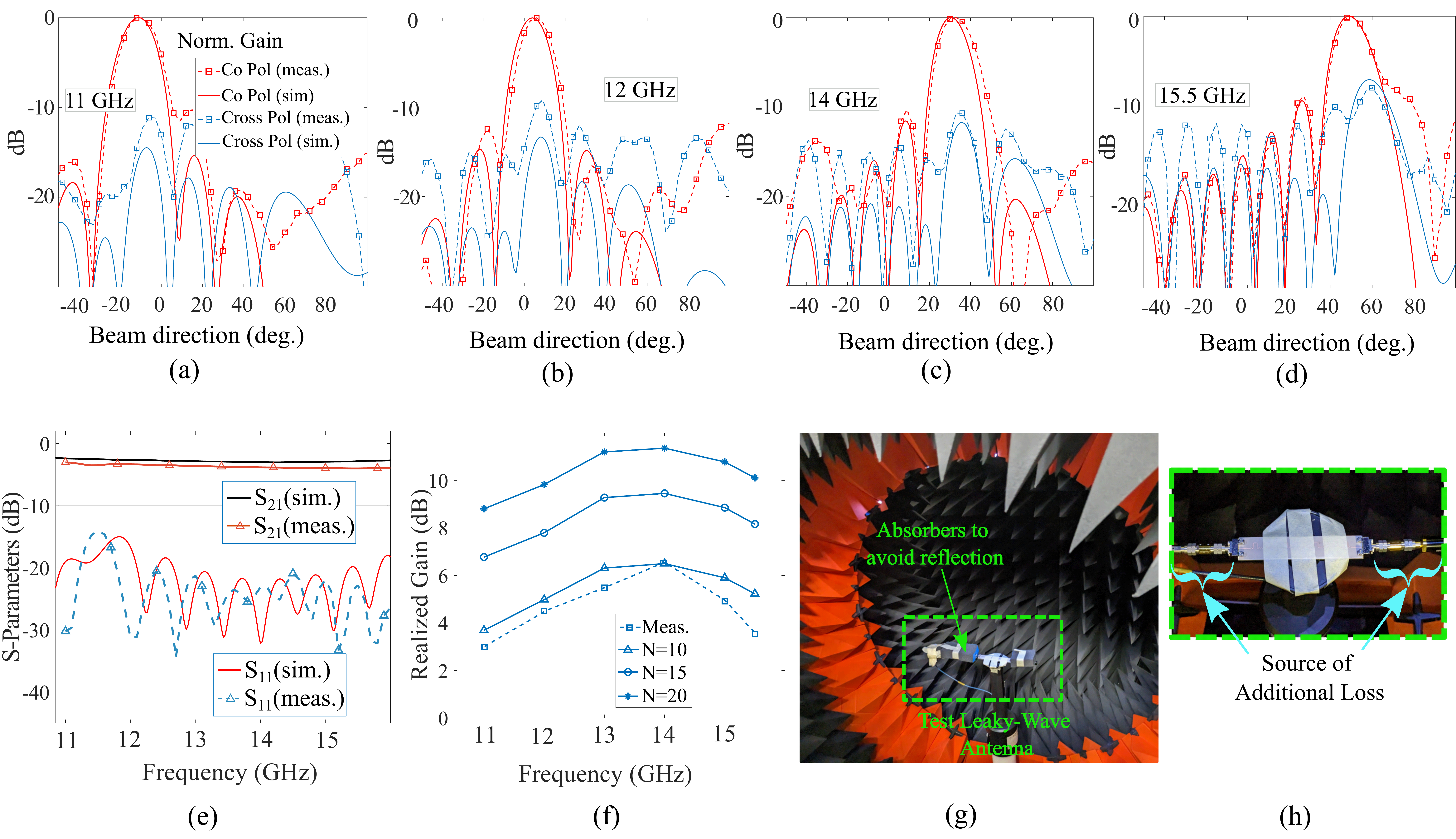}}
\caption{Measurements in the Ku-band where the antenna is linearly polarized. Measured and simulated normalized radiation pattern of the fabricated antenna in the H-plane at (a) \SI{11}{GHz}, (b) \SI{12}{GHz}, (c) \SI{14}{GHz}, (d) \SI{15.5}{GHz}. (e)~{Simulated and measured S-parameters.} (f)~Realized gain in Ku-band with different number of unit cells obtained through full-wave simulations. (g)~Experimental setup for the measurement in Ku-band. (h) The additional connectors required to connect the antenna to the ports of VNA.}
\label{KuBandMeasure}
\end{figure*}

\section{Dual-Band Leaky-Wave Antenna}
\label{sec:measurement}
A 10 unit cell dual band \gls{lwa}, having the layout of Fig.~\ref{thetavsfreq}(b) and the optimised dimensions listed in Table~\ref{3FingUnitCellDimensions_Table} was fabricated. The antenna is based on Rogers 3003 ($\varepsilon_r = 3.0$ and $\tan\delta=~0.001$) with a substrate thickness of \SI{0.254}{\milli\meter}.

\captionsetup[table]{name=TABLE,labelsep=newline,textfont=sc}
\renewcommand{\arraystretch}{1.5}
\begin{table*}[t!]
\caption{Comparison with other scanning \glspl{lwa}\label{CompareTable} operating in similar frequency ranges}
\centering
\begin{tabular}[t]{l>{\centering}p{0.20\linewidth}>{\centering\arraybackslash}p{0.14\linewidth}>{\centering\arraybackslash}p{0.115\linewidth}>{\centering\arraybackslash}p{0.08\linewidth}>{\centering\arraybackslash}p{0.11\linewidth}p{0.055\linewidth}}
\toprule
\textbf{Ref}&\textbf{Antenna Type}&\textbf{Frequency BW }&\textbf{Scanning Range}&\textbf{Realized Gain (dB)}&\textbf{Rel. Permittivity ($\varepsilon_r$)} & \textbf{Pol.} \\
\midrule
\cite{rahmani_backward_2017}& Microstrip with metallized vias & \SIrange{20}{29}{\giga\hertz} & \SI{95}{\degree} (\SI{-50}{\degree} to \SI{45}{\degree}) & Approx. 10 & 6.15 & Circular \\
\cite{zhao_circularly_2022} & Microstrip with metallized vias & \SIrange{9.8}{15}{\giga\hertz} & \SI{72}{\degree} (\SI{-52}{\degree} to \SI{20}{\degree}) & 10.0--15.1 & 2.2 & Circular \\
\cite{geng_ka-band_2022}& SIW & \SIrange{26}{43}{\giga\hertz} & \SI{70}{\degree} (\SI{-40}{\degree} to \SI{30}{\degree}) & 13.4--15 & 3.0 & Linear \\ 
\cite{otto_circular_2014}& SIW & \SIrange{23}{25}{\giga\hertz} & \SI{30}{\degree} (\SI{-15}{\degree} to \SI{15}{\degree}) & 15 & 3.66 & Circular \\
\cite{lyu_periodic_2017}& SIW & \SIrange{10}{14}{\giga\hertz} & \SI{65}{\degree} (\SI{-40}{\degree} to \SI{25}{\degree}) & 5.1--13 & 3.0 & Circular \\
\cite{vadher_higher_2023} & Microstrip & \SIrange{20.6}{24.6}{\giga\hertz} & \SI{85}{\degree} (\SI{-40}{\degree} to \SI{45}{\degree}) & 9.8--13 & 3.0 & Linear \\
This Work -- K-band & Microstrip & \SIrange{19.4}{27.5}{\giga\hertz} & \SI{72}{\degree} (\SI{-42}{\degree} to \SI{30}{\degree}) & 7--13 & 3.0 & Circular \\
This Work -- Ku-band         &     Microstrip       & \SIrange{11}{15.5}{\giga\hertz} & \SI{75}{\degree}(\SI{-15}{\degree} to \SI{60}{\degree}) & {3.1--6}& 3.0 & Linear \\
\bottomrule
\end{tabular}
\end{table*}
\subsection{Feed design}
A Bulgin end-launch connector (\SI{2.4}{\milli\meter}) with the output impedance $Z_0=50\ \Omega$ is used to feed the antenna. The $Z_\text{{S}}$ of the antenna is around $60\ \Omega$ throughout the operating frequency range as shown in Fig.~\ref{Z_bloch_OSBRemoval}. Hence a matching section is added to match the impedance depicted in Fig.~\ref{FeedStructure}.

\subsection{Circularly polarized radiation in K-band}
In this band, the antenna is right-handed circularly polarized and radiation is due to spatial harmonic of $n=-2$. The measurements are performed at the millimeter-wave test facility CAMILL at Institut
d’Électronique et de Télécommunications de Rennes (IETR). The simulated and measured radiation pattern in the azimuth plane are shown in Fig.~\ref{KBandMeasure}(a--d) for the \SIrange{19.4}{27.5}{\GHz} band. The measured radiation pattern at $f=\SI{19.4}{\giga\hertz}$ shows higher side-lobes compared to simulations. This is due to the fact that there is reflection from the connector due to a very tilted beam in the H-plane. The 3-\SI{}{\dB} beamwidth of the antenna is about $10\degree$ throughout the operational range. The length of the antenna remains compact at 7.67$\lambda_0${($=c/f_\textrm{b}$)}.

 Measured and simulated axial ratio are compared in Fig.~\ref{KBandMeasure}(e). The realized gain plot and directivity of the reconfigurable fan-beam radiation pattern in the H-plane is also shown in Fig.~\ref{KBandMeasure}(e). The measured S-parameters show that the reflection coefficient remains below \SI{-10}{\dB} for the operational range. {The simulated radiation efficiency ($\mathrm{\eta_{{rad}}}$) of the \gls{lwa} (as defined in \cite{jackson_leaky-wave_2008}) in this band varies from 0.5 to 0.75. The measured total radiation efficiency ($\mathrm{\eta_{{tot}}}$) varies from 0.45 to 0.72 as shown in Fig.~\ref{KBandMeasure}(f). The total radiation efficiency contains the effect of mismatch loss due to matching circuit\cite{nguyen_highly_2009}. The total radiation efficiency of the antenna is obtained by dividing the measured realized gain with directivity \cite{massoni_increasing_2019, sarkar_60_2020}}. 

\subsection{Linearly polarized radiation in Ku-band}
The antenna is linearly polarized in the Ku-band (\SIrange{11}{15.5}{\GHz}) and the radiation occurs due to the $n = -1$ spatial harmonic. The simulated and measured realized gain are shown in Fig.~\ref{KuBandMeasure}(a)--(d). The antenna scanning range is of \SIrange{-15}{60}{\degree}. The 3-\SI{}{\dB} beamwidth of the antenna is less than \SI{17}{\degree} throughout the operational range. The measured reflection coefficient is below \SI{-10}{\dB} in the operating frequency range as shown in Fig.~\ref{KuBandMeasure}(e).

Measurement is performed using MVG Starlab measurement system at IETR, as shown in Fig.\ref{KuBandMeasure}(g). The antenna is fixed on the platform using tapes to avoid displacement. To avoid the elevated side-lobe levels at high tilted angles absorbers are added at the either end.

It has to be noted that the antenna has been designed to operate in K-band and Ku-band, with a special attention to the performance in K-band. As a consequence, the gain in Ku-band is limited to {\SI{6}{\dB}}. However, the realized gain can be improved by adding unit cells to the \gls{lwa} as shown in Fig.~\ref{KuBandMeasure}(f). {Another approach is to increase the thickness of the dielectric substrate (Rogers 3003) on which the \gls{lwa} is designed to improve the realized gain. This approach is discussed in detail in the subsequent subsection (Section \ref{subsec:ImpKuBand})}. It is also important to notice that the realized gain in the Ku-band of the antenna is lower by \SIrange[range-units=single, range-phrase=--]{1}{2}{\dB} in the measurements than simulations as shown in Fig.~\ref{KuBandMeasure}(f). This is the effect of the connectors required to properly connect the antenna to the ports of \gls{vna} as depicted in Fig.~\ref{KuBandMeasure}(h).

\subsection{Literature comparison}
 Table \ref{CompareTable} shows the comparison with the antennas operating in similar frequency ranges along with fabrication technology. Circularly polarized \glspl{lwa} operating in the range \SIrange{9.8}{15}{\giga\hertz} and \SIrange{20}{29}{\giga\hertz} based on microstrip are reported in \cite{zhao_circularly_2022, rahmani_backward_2017}. Both designs require via-holes, thus increasing the complexity for fabrication compared to the proposed design. The works in \cite{lyu_periodic_2018, vadher_higher_2023} report \gls{lwa} based on microstrip operating in similar mm-Wave range however they exhibit linear polarization. \Gls{crlh} based structures such as the ones reported in \cite{dong_composite_2009, sabahi_compact_2018,agarwal_multilayered_2021} are extremely sensitive to dimensions of the slot making them difficult to fabricate with precision at high frequencies.
 
\begin{figure}[!t]
\centerline{\includegraphics[width=0.42\textwidth]{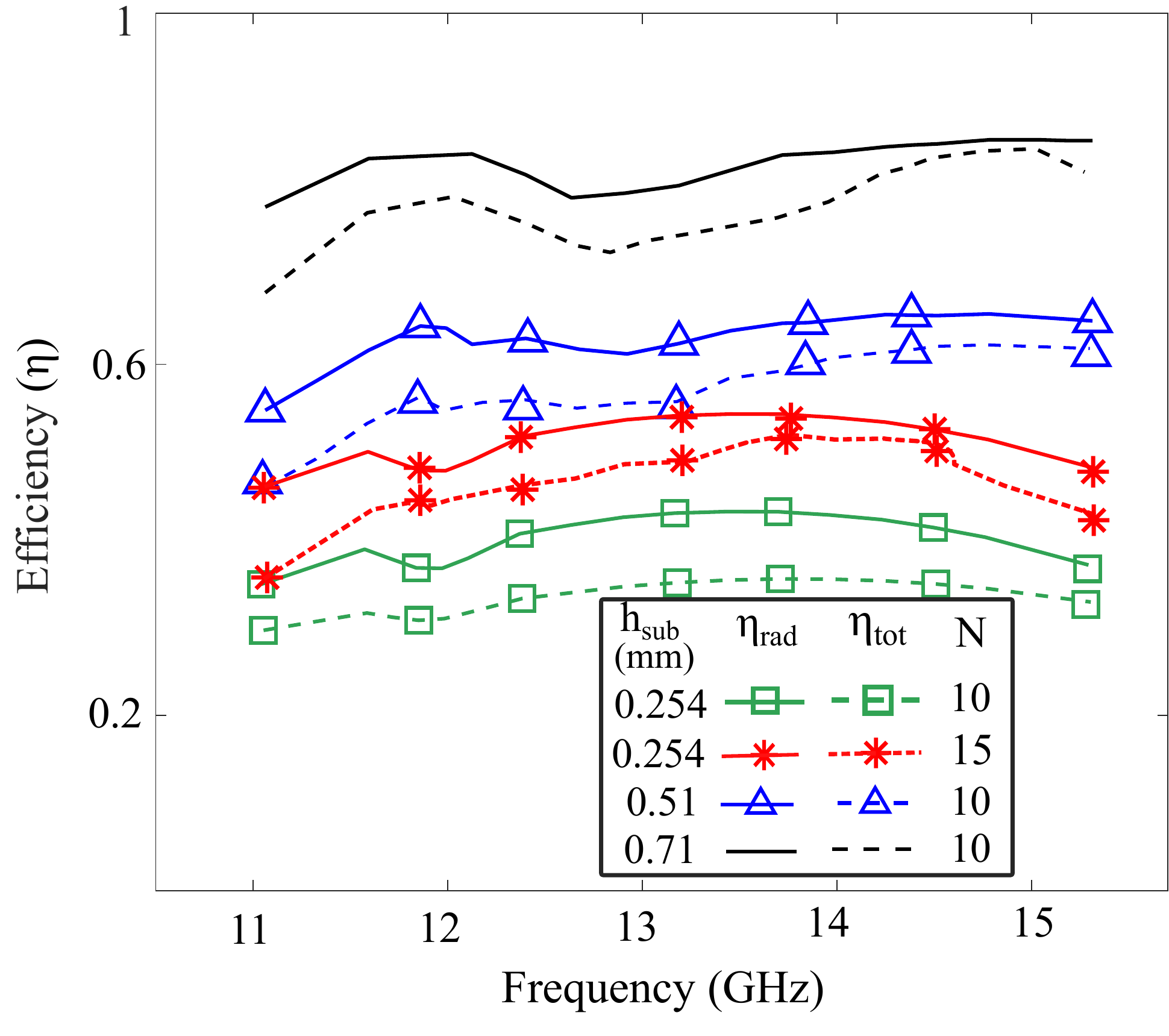}}
\caption{{Efficiency as a function of frequency in Ku-band for different dielectric (Rogers 3003) thickness ($\mathrm{h_{sub}}$) of the LWA keeping all the other dimensions unchanged.}}
\label{EffLowerBand}
\end{figure}

\begin{figure}[!t]
\centerline{\includegraphics[width=0.405\textwidth]{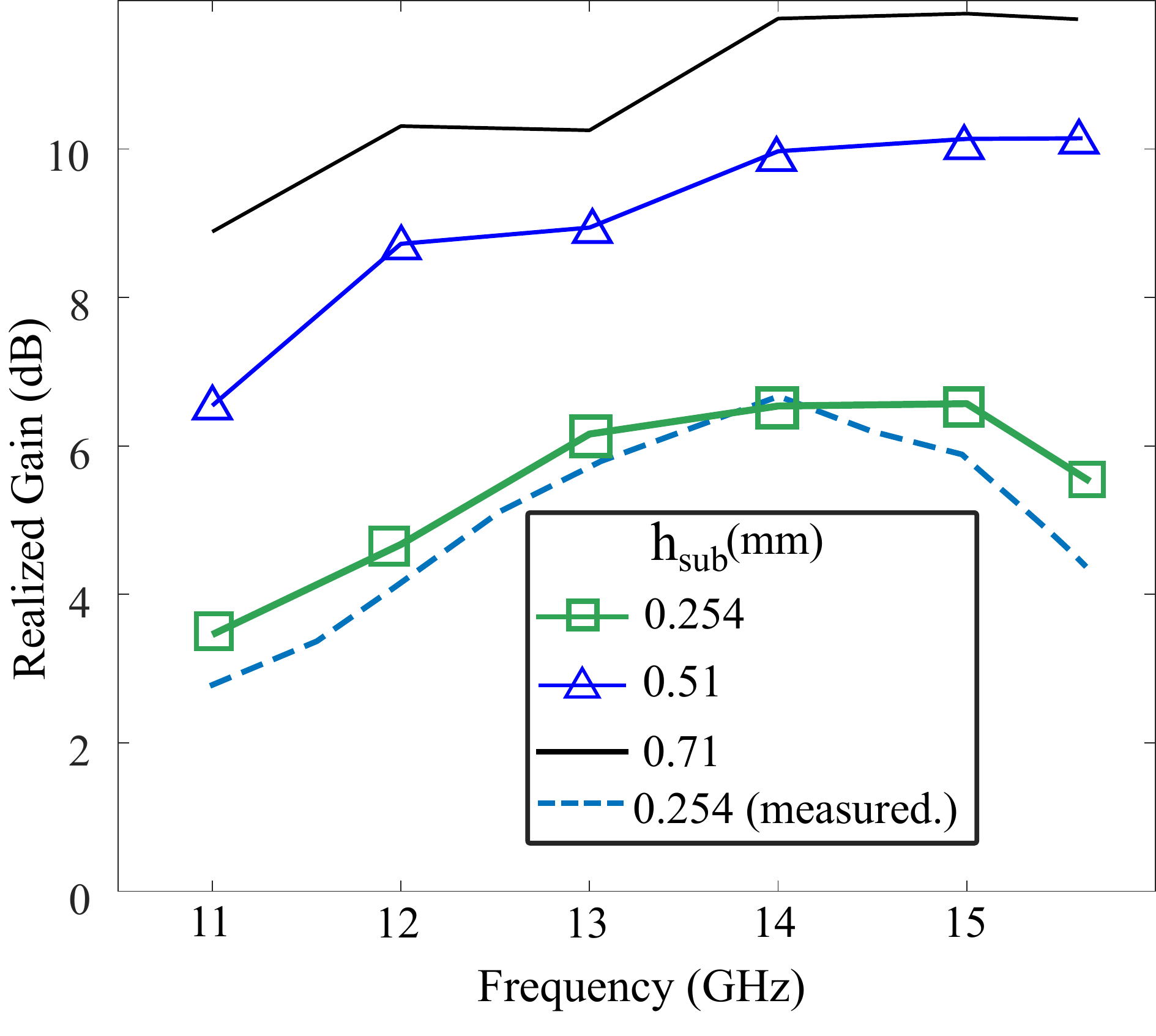}}
\caption{{Increase in realized gain in Ku-band for different dielectric thickness for an LWA comprising of $\mathrm{N}=10$ unit cells.}}
\label{GainIncLowerBand}
\end{figure}

 Meanwhile, the proposed antenna achieves large beam scanning angles with simple meandering microstrips, while maintaining circular polarization. The design can be scaled at lower or higher frequencies very easily without the increase in complexity of fabrication.
{\subsection{Improving the low gain and low efficiency of the antenna in the Ku-band}\label{subsec:ImpKuBand}}
{The efficiency and gain of the prototyped \gls{lwa} in the Ku-band is shown in Fig.~\ref{EffLowerBand} and Fig.~\ref{GainIncLowerBand} respectively (corresponding to $\mathrm{N}=$ 10 and substrate thickness of $\mathrm{h_{sub}}=$ \SI{0.254}{\milli\meter}). To improve the gain of the antenna in the Ku-band, more than $\mathrm{N}=10$ unit cells can be cascaded as suggested in the Subsection \ref{subsec:scanningrate} and Fig.~\ref{KuBandMeasure}(f). However, the efficiency of the \gls{lwa} remains low even with a higher number of unit cells in the Ku-band as shown in Fig.~\ref{EffLowerBand} ($\mathrm{N}=$ 15 and $\mathrm{h_{sub}}=$ \SI{0.254}{\milli\meter}). This indicates that the prototyped \gls{lwa} is inherently a poor radiator in the Ku-band.}

{
To enhance the efficiency and overall performance of the antenna without increasing the number of unit cells in the original \gls{lwa} (i.e., $\mathrm{N = 10}$), one approach is to increase the dielectric substrate thickness while keeping all other dimensions unchanged. This reduces the coupling of the electric fields from the microstrip to the ground while increasing the fringing fields \cite{pozar_microwave_2011} resulting in increased radiation from the mitred corners. This makes the \gls{lwa} a better radiator in the Ku-band (and K-band as well). Fig.~\ref{EffLowerBand} shows the improvement in efficiency when the thickness of substrate is increased while cascading the same number of unit cells as the prototyped \gls{lwa} ($\mathrm{N}=10$). The realized gain of the \gls{lwa} also increases as shown in the Fig.~\ref{GainIncLowerBand} for the same physical length ($=\mathrm{N}p$). Since the increase in thickness is essentially increasing the aperture efficiency resulting in lower aperture length of the antenna, this leads to an increase in the 3-\SI{}{\dB} beamwidth of the \gls{lwa} \cite{jackson_leaky-wave_2008}. For the substrate thickness of \SI{0.254}{\milli\meter}, \SI{0.51}{\milli\meter} and \SI{0.71}{\milli\meter} the 3-\SI{}{\dB} beamwidth is \SI{17}{\degree}, \SI{18.7}{\degree} and \SI{19.5}{\degree} respectively. The antenna retains all the other properties, including the beam scanning with frequency, linear polarization in the Ku-band and circular polarization in the K-band.}

{
Even though the substrate thickness of \SI{0.254}{\milli\meter} is responsible for a lower efficiency in Ku-band, this thickness was chosen for the prototype to maximise the bendability of the antenna for on-body applications. }

\bigskip

\section{Conclusion}
\label{sec:conclusion}
This paper presents a novel approach to designing a compact and single-layer circularly polarized \gls{lwa}. The approach is based on utilizing the time delay of microstrip line intervals to achieve circular polarization. The frequency scanning \gls{lwa} operates as circularly polarized antenna in the K-band and as linearly polarized antenna in the Ku-band. This is possible by the use of spatial harmonics $n=-2$ and $n=-1$, respectively. To improve the band-separation and reduce the side-lobes due to unwanted harmonics in the radiation zone, additional meanders have been introduced. The axial ratio remains below \SI{3}{\decibel} over large frequency range (\SIrange{19.4}{27.5}{\giga\hertz}) in the intended band for circular polarization. This ensures large scanning range from \SI{72}{\degree} (\SI{-42}{\degree} to \SI{30}{\degree}). A novel technique to remove \gls{osb} is also explained in the work in order to have continuous scanning through broadside direction. 

The proposed antenna does not make use of metallized vias hence significantly reducing the complexity of fabrication while maintaining compact size (7.67$\lambda_0$). The antenna is manufactured on a flexible substrate of Rogers 3003, making it suitable for flexible and conformal purposes at mm-wave frequencies.


\bibliographystyle{IEEEtran1}
\bibliography{MyLibrary_17_04_23}

\end{document}